\documentclass[aps,prb,twocolumn,groupedaddress,showpacs,amssymb,amsmath]{revtex4-1}
\usepackage{braket}
\usepackage{graphicx}
\usepackage{color}
\usepackage{pgf}
\usepackage{epstopdf}
\usepackage{amsmath}
\usepackage{float}
\begin{document}
\title{Entanglement entropy and entanglement spectrum of Bi$_{1-x}$Sb$_{x}$ (111) bilayers}
\author{Marta Brzezi\'{n}ska}
\email{marta.a.brzezinska@pwr.edu.pl}
\author{Maciej Bieniek}
\author{Tomasz Wo\'{z}niak} 
\author{Pawe\l{} Potasz}
\author{Arkadiusz W\'{o}js}
\affiliation{Department of Theoretical Physics, Faculty of Fundamental Problems of Technology, Wroc\l{}aw University of Science and Technology, 50-370 Wroc\l{}aw, Poland}
\date{\today}

\begin{abstract}
We investigate topological properties of Bi$_{1-x}$Sb$_{x}$ bilayers in the (111) plane using entanglement measures. Electronic structures are studied using multi-orbital tight-binding model, with structural stability confirmed through first-principles calculations. Topologically non-trivial nature of Bi bilayer is proved by the presence of spectral flow in the entanglement spectrum. We consider topological phase transitions driven by a composition change $x$, an applied external electric field in pure Bi, and strain in pure Sb. Composition- and strain-induced phase transitions reveal a finite discontinuity in the entanglement entropy. However, this quantity remains a continuous function of electric field strength, but shows a discontinuity in the first derivative. We relate the difference in behavior of the entanglement entropy to breaking of the inversion symmetry in the last case.
\end{abstract}
\maketitle

\section{Introduction}
Exploring novel phases of quantum matter has gathered interest due to promising applications in the fields of spintronics and quantum computation \cite{Hasan:rev, Zhang:rev, QComp}. Considerable effort has been devoted to searching candidate materials for topological insulators (TIs), a class of band insulators hosting gapless edge states and described by a $\mathbb{Z}_2$ topological invariant \cite{KaneMele:QSHE, Schnyder:class, FuKane:Z2}. Boundary modes lead to two spin-polarized counter-propagating currents, which are immune to the backscattering in presence of non-magnetic disorder \cite{Hasan:rev, Zhang:rev}. For quantum dot geometry, orbital magnetization resulting from edge state circulation has also been shown to exhibit similar robustness \cite{potasz}. Essential ingredients for quantum spin Hall (QSH) phase are the time-reversal symmetry and spin-orbit coupling (SOC). Strong SOC is characteristic of heavy elements and opens the non-trivial bulk band gap. Proposals of realizing the QSH insulator in two dimensions (2D) \cite{KaneMele:QSHE, QSHE:Bernevig, BHZ, QSHE:Zhang} were followed by successful experimental observations, ranging from thickness-tunable quantum wells to honeycomb-like systems based on group-IV and group-V elements \cite{ren2016topological}, and recently as thin films of insulators protected by crystalline symmetries \cite{TCI, TCI2}. 

In particular, theoretical predictions have shown that a single Bi bilayer in the (111) plane manifests helical edge modes propagating in opposite directions  \cite{Murakami:BiQSH, stable:bi111, singlebi111, ElTPbi} and subsequently it was confirmed by scanning tunneling microscopy measurements \cite{drozdov, kawakami2015one, spatialandenergy}. Moreover, protected one-dimensional edge states were detected in Bi$_2$Se$_3$ thin films \cite{BiSe2d, Zhang:1} and at the interface between heterostructures Bi(111)/Bi$_2$Te$_3$ \cite{BiTe3inter}. Robustness of topological properties of Bi(111) were lately disscused in Refs. \cite{stability:bi111, Bi:robust, edgeen}. Extensive studies have been made for few-layer Bi and Sb systems \cite{koroteev, pan2015realization, bian2016engineering, PhysRevMaterials.1.014002}. Pure Bi(111) films up to 8 bilayers thickness were shown to exhibit stable QSH phase \cite{stable:bi111}. In the case of Sb(111), films with less than four bilayers are expected to be topologically trivial \cite{Sb:triv}. To induce transition between trivial and QSH insulating phases, appropriate structure modifications were proposed, including spin-orbit coupling variation in Bi \cite{stability:bi111, BiAPPA}, strain in Sb \cite{jin2015quantum, Sb:nontriv}, effect of substrate \cite{singlebi111} and perpendicular electric field for strained Sb \cite{wang2014topological}.

Over the years, quantum entanglement has been recognized as a valuable tool to identify topological properties of the systems. Subtle non-local correlations inherent in topological states can be captured by bipartite entanglement measures \cite{Eisert:EE, Ent:rev1}. Given a composite system $AB$ in a pure state described by $\ket{\psi_{AB}} \in \mathcal{H}_A \otimes \mathcal{H}_B$, the reduced density matrix corresponding to the subsystem $A$ can be obtained by tracing out degrees of freedom related to the subsystem $B$, $\rho_A = \textnormal{Tr}_B \ket{\psi_{AB}} \bra{\psi_{AB}}$. Commonly used in the context of quantum information is the von Neumann entanglement entropy, defined as $S_A = - \textnormal{Tr} \rho_A \log \rho_A$. Li and Haldane \cite{LiHaldane:ES} suggested that the Schmidt decomposition of quantum many-body wave function can be regarded as a unique signature of topological phase and provides more information about the system than a single value of $S_A$. Several non-interacting models describing clean \cite{ENT:super1, ENT:super1, Fid:ES, Bernevig:class, Vish:inv, Fritz:ES, PhysRevB.82.085106, Alex:CM, Hughes:inv} as well as disordered systems \cite{Bernevig:Disord, Hughes:Disord1, Hughes:Disord2} have been already analyzed through entanglement spectrum. 

In this work, we analyze topological properties of Bi$_{1-x}$Sb$_{x}$ bilayer through entanglement measures. We examine electronic properties and structural stability of Bi$_{1-x}$Sb$_{x}$ bilayers, which is achieved by employing multi-orbital tight-binding model and density functional theory (DFT) calculations. Both infinite and semi-infinite systems are considered. We investigate entanglement entropy and entanglement spectrum for free-fermion systems and show that they can be treated as viable tools to characterize topological properties of realistic models of topological insulators. Furthermore, topological phase transitions as a result of composition change $x$, perpendicular electric field in pure Bi, and strain in pure Sb are studied.

\section{\label{sec:Meth}Methodology}
\subsection{\label{sec:Mod}Tight-binding model}
\begin{figure}[H]
	\centerline{\includegraphics[width=0.85\columnwidth]{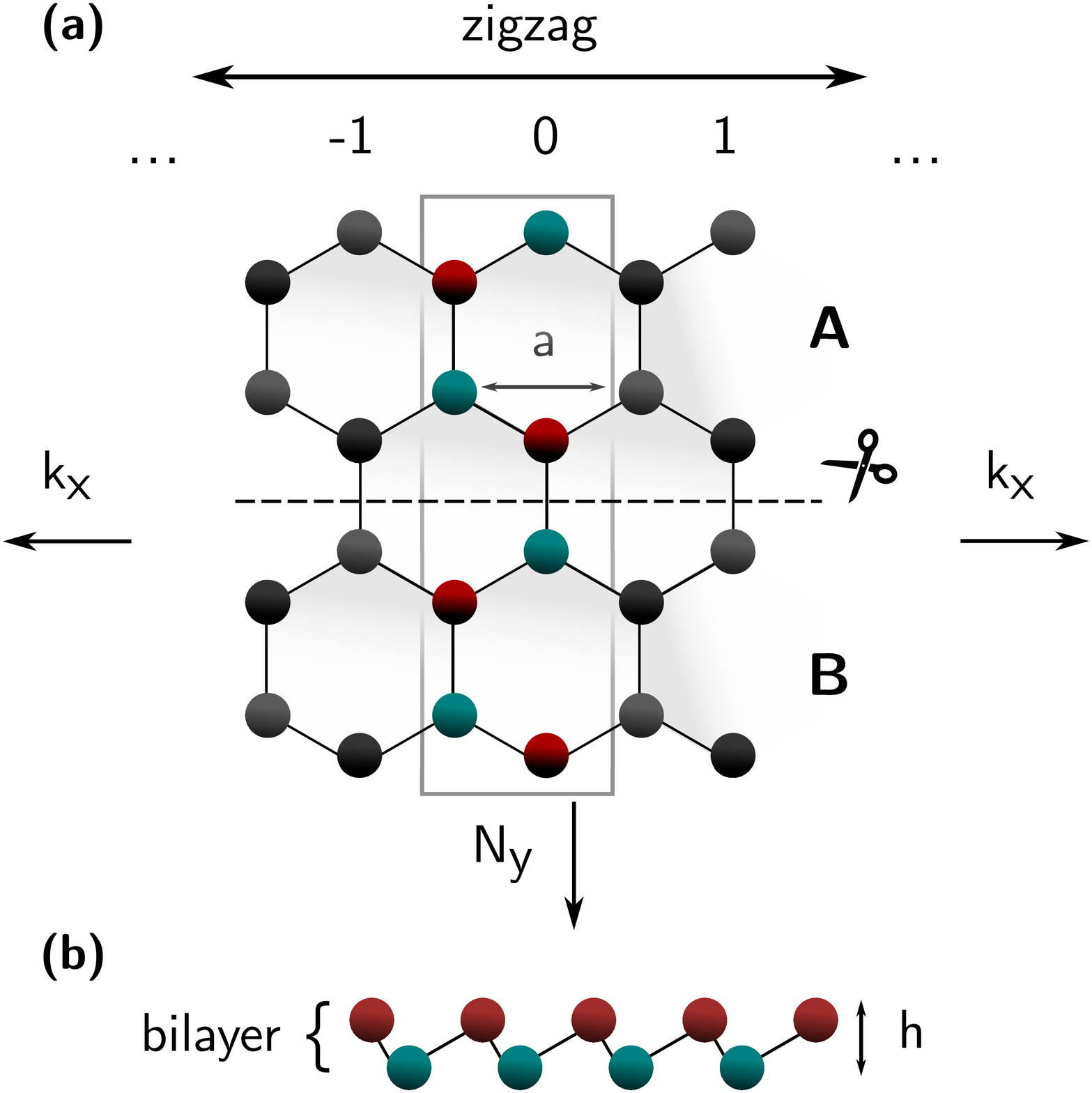}}
	\caption{(Color online) Lattice structure of Bi and Sb bilayer. (a) Top view of a ribbon with a zigzag edge termination of width $N_y$ and periodic boundary conditions in $x$ direction. Four atoms are within the unit cell of a ribbon. $a$ the is lattice constant, $0$ labels a unit cell, while $-1$ and $1$ denote left and right neighboring cells, respectively. Red and green colors distinguish two sublattices forming a honeycomb lattice. The system can be divided into two halves $A$ and $B$, with a cut marked by a dashed line, in order to evaluate entanglement measures. (b) Side view of a bilayer. $h$ corresponds to the bilayer thickness.}
	\label{fig:latt_struct}  
\end{figure}

Single sheet of Bi and Sb rhombohedral crystal structures in the (111) direction can be seen as a buckled honeycomb lattice and is illustrated in Fig. \ref{fig:latt_struct}. We follow a widely used terminology calling a single sheet by bilayer \cite{Murakami:BiQSH, drozdov}, as it consists of two sublattices forming two layers spatially separated by a distance $h$.

To study electronic properties of ultrathin films, we employ sp$^3$ tight-binding model developed in Ref. \cite{Liu:Allen} for bulk bismuth and antimony, neglecting hoppings between bilayers as proposed in Ref. \cite{Murakami:BiQSH}. Interatomic hoppings are parametrized within the Slater-Koster approach \cite{Slater:Koster} and atomic spin-orbit coupling $\lambda \vec{L}\cdot\vec{S}$. The Hamiltonian of the model reads
\begin{equation}
\begin{aligned}
&H_{TB} = \sum_{i, \alpha,\sigma}E_{i\alpha}c^\dagger_{i\alpha\sigma}c_{i\alpha\sigma} +\sum_{i,\alpha,\sigma}E_{Field}^{R,G}c^\dagger_{i\alpha\sigma}c_{i\alpha\sigma}  \\
& + \left[ \sum_{i,j,\alpha,\alpha',\sigma}V_{\alpha\alpha'}c^\dagger_{i\alpha\sigma}c_{j\alpha'\sigma} +\frac{\lambda}{3}\sum_{i}(c^\dagger_{iz\downarrow}c_{ix\uparrow} - c^\dagger_{iz\uparrow}c_{ix\downarrow} \right. \\
& \left. +ic^\dagger_{iz\uparrow}c_{iy\downarrow}+ic^\dagger_{iz\downarrow}c_{iy\uparrow} +ic^\dagger_{ix\downarrow}c_{iy\downarrow}-ic^\dagger_{ix\uparrow}c_{iy\uparrow})+ H. c. \right] ,
\end{aligned}
\label{eq:hamiltonian}
\end{equation}
where $i, j$ are lattice indices, $\alpha = \lbrace s, p_x, p_y, p_z \rbrace$ labels orbitals an $\sigma= \lbrace \uparrow,\downarrow \rbrace$ denotes spins. $E_{i}$ corresponds to the on-site energies and $V_{\alpha\alpha'}$ are the hopping integrals. $E_{Field}^{R} =-E_{Field}^{G} =E_{Field}$ is a perpendicular electric field on $i\in R$ and $i\in G$ sites of two sublattices in a lattice indicated by red and green color in Fig. \ref{fig:latt_struct}. The last term is the SOC with a strength $\lambda$. According to Chadi \cite{Chadi}, 1/3 factor is introduced to renormalized atomic SOC in order to obtain correct SOC splitting of the valence band. 

\begin{table}
\centering
\begin{tabular}{|c|c|c||c|c|c|}
\hline 
Parameter & Bi & Sb & Parameter  & Bi & Sb \\ 
 (eV) &  &  & (eV)  &  &  \\ \hline
E$_{s}$ & -10.906 & -10.068 & V'$_{sp\sigma}$ & 0.433 & 0.478 \\ 
E$_{p}$ & -0.486 & -0.926 & V'$_{pp\sigma}$  & 1.396 & 1.418  \\
V$_{ss\pi}$ & -0.608 & -0.694 & V$'_{pp\pi}$ & -0.344 & -0.393 \\
V$_{sp\sigma}$ & 1.320 & 1.554 & V''$_{ss\sigma}$ & 0 & 0 \\
V$_{pp\sigma}$ & 1.854 & 2.342 & V''$_{sp\sigma}$ & 0 & 0\\
V$_{pp\pi}$& -0.600 & -0.582 & V''$_{pp\sigma}$ & 0.156 & 0.352 \\
V'$_{ss\sigma}$& -0.384 & -0.366 & V''$_{pp\pi}$ & 0 & 0\\
$\lambda$ & 1.5 & 0.6 &   & & \\
\hline
a ($\textnormal{\AA}$) & 4.53 & 4.30 & h ($\textnormal{\AA}$) & 1.58$^{*}$ & 1.64$^{**}$ \\
d$_1$ ($\textnormal{\AA}$) & 3.062 & 2.902 &  &  &\\
\hline
\end{tabular} 
\caption{Tight-binding parameters for Bi and Sb taken from Refs. \cite{Liu:Allen}, \cite{stable:bi111}$^*$, \cite{Sb:nontriv}$^{**}$. $d_1$ denotes nearest-neighbor distance between sites in a honeycomb lattice.}
\label{tab:TB}
\end{table}

Similarities between Bi and Sb crystals can be seen by looking at Slater-Koster parameters listed in Table I, taken from Ref. \cite{Liu:Allen}. Most of the parameters differ by less than $15\%$. The only significant change is in the spin-orbit coupling constant $\lambda$, $2.5$ times larger in bismuth. Thus, a transition from QSH insulating phase to trivial insulator with increasing $x$ in Bi$_{1-x}$Sb$_x$ is related, in general, to a decrease of spin-orbit coupling constant.

\subsection{\label{sec:DFT}DFT stability calculations of Bi$_{1-x}$Sb$_x$ alloys}
We investigate stability of Bi$_{1-x}$Sb$_x$ alloys with different composition $x$. DFT calculations have been performed in ABINIT software \cite{DFT1} using the LDA exchange-correlation functional. The Bi and Sb atoms were modelled by fully relativistic PAW sets. The plane wave basis cut-off was 20 Ha and the Monkhorst-Pack k-point grid was set to 10x10x1. The structures were fully relaxed until the interatomic forces were lower than 10$^{-8}$ Ha/Bohr. The phonon calculations were performed in Phonopy software \cite{DFT2} which implements Parlinski-Li-Kawazoe method that is based on the supercell approach with the finite displacement method \cite{DFT3}. 3x3x1 supercells were used for modelling of Bi$_{1-x}$Sb$_x$ mixed crystals. Phonon band structures were calculated along $\Gamma$-M-K-$\Gamma$ path in the reciprocal space. The supercells of given size are commensurate with all the q-vectors of the high symmetry points in hexagonal Brillouin zone (BZ) \cite{DFT4}. 

\begin{figure}[H]
\centering
\includegraphics[width=0.85\columnwidth]{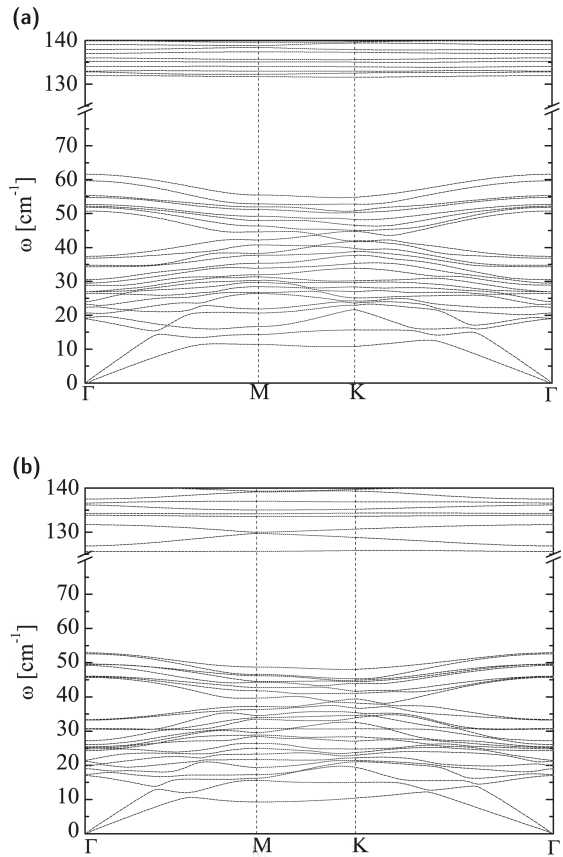}
\caption{Phonon dispersion of (a) Bi$_{0.28}$Sb$_{0.72}$ and (b) Bi$_{0.72}$Sb$_{0.28}$ mixed crystals. Lack of modes with imaginary frequencies indicates stability of the crystals.}
\label{fig:DFT}
\end{figure}

We have modeled Bi$_{0.72}$Sb$_{0.28}$ and Bi$_{0.28}$Sb$_{0.72}$ mixed crystals with phonon dispersions shown in Fig. \ref{fig:DFT}. The multiple phonon branches are the consequence of folding of the phonon states in the supercell first Brillouin zone. We observe no imaginary phonon modes, which confirms the structural stability. Very recently, Bi$_{0.5}$Sb$_{0.5}$ has also been found stable by employing phonon calculations \cite{DFT:BiSb}.

\subsection{\label{sec:EntMeas}Entanglement measures for free-fermion systems}
Quantum correlations between parts of the composite system can be studied quantitatively by means of entanglement measures \cite{Plenio:2007zz}. Suppose the system is bipartited into two equal spatial parts, $A$ and $B$, as in Fig \ref{fig:latt_struct}(a). Information about the part $A$ is encoded into reduced density matrix $\rho_A$. In particular, $\rho_A$ can be represented as $\rho_A = e^{ - H_A } /Z_A$, with $Z_A = \textnormal{Tr} \left( e^{ - H_A } \right)$ being the partition function. The matrix $H_A$ is a logarithm of thermal reduced density matrix at the temperature $T = 1$, $H_A = - \log \rho_A$, and can be identified as the entanglement Hamiltonian. Therefore, the entanglement spectrum (ES) is defined as a set of eigenvalues of $H_A$ denoted by $\lbrace \xi \rbrace$. 

Entanglement measures for free fermionic lattice systems can be computed from the two-point correlation function restricted to the subsystem \cite{Peschel}
\begin{equation}
C^{\alpha \beta}_{ij}= \textnormal{Tr} \left( \rho_A c^{\dagger}_{i\alpha} c_{j\beta} \right),
\end{equation}
where $i, j$ are lattice indices within the subsystem $A$ and $\alpha, \beta$ label orbitals or spins. If system is translationally-invariant, the Hamiltonian $H$ can be written in the momentum space with the many-body ground state in a form $\ket{GS} = \prod_{n, k}  c^{\dagger}_{n k} \ket{0}$ with $k$ being the conserved momentum and $n$ running over the occupied single-particle Bloch states. Hence, the correlation matrix can be evaluated for each $k$-point separately via formula $C^{\alpha \beta}_{ij} (k) = \braket{GS | c^{\dagger}_{i\alpha k} c_{j \beta k} | GS}$. $C(k)$ is a Hermitian matrix and can be regarded as a spectrally flattened physical Hamiltonian, with eigenvalues (denoted by $\lbrace \zeta_k \rbrace$) falling between $0$ and $1$. Most of the eigenvalues in the spectrum of $C(k)$ lie exponentially close to either 1 or 0, depending whether bulk states are fully localized in the subsystem $A$ or $B$, respectively, and do not contribute to the entanglement entropy. However, states crossing the partition boundary give rise to non-zero entanglement entropy. If the Hamiltonian describes a topologically non-trivial phase, $C(k)$ will reveal the spectral flow associated with continuous set of intermediate eigenvalues \cite{Hughes:inv, Vish:inv}. A relation between $\lbrace \zeta_k \rbrace$ and the spectrum of $H_A(k)$ labeled by $\lbrace \xi_k \rbrace$ is following
\begin{equation}
\zeta_k = \left( 1+e^{\xi_k} \right)^{-1}.
\end{equation}
Due to this one-to-one correspondence, we refer to eigenvalues of the correlation matrix as the single-particle entanglement spectrum, which is a conventional practice in the literature \cite{Alex:CM, Vish:inv, Hughes:inv}. Entanglement entropy is given by
\begin{equation}
S_A = - \sum_a \left( \zeta_a \log \zeta_a + \left( 1 - \zeta_a \right) \log \left( 1 - \zeta_a \right) \right),
\label{eq:entCij}
\end{equation}
where $a$ is index running over all eigenvalues of $C$. $S_A$ can be calculated by summing over the entanglement entropy for each $k$-point with a normalization factor being the number of unit cells $S_{A} = 1 / N_y \sum_k S_A(k)$ \cite{Ryu:EE}.

It was also shown that the trace of $C$ corresponding to the subsystem $A$ called trace index is equivalent to the topological invariants for AII and A symmetry classes \cite{Alex:CM}. Counting the discontinuities in the trace index provides a method to extract $\mathbb{Z}_2$ invariant. Physical edge states in the bulk gap that cross the Fermi level translate into discontinuities in the $\textnormal{Tr} \, C (k)$, thus trace index would not be applicable in the absence of boundary modes.


\section{\label{sec:PureBiSb} Entanglement spectra of pure Bi and Sb}
\subsection{Bi and Sb infinite bilayers}
\begin{figure}[H]
	\centerline{\includegraphics[width=0.92\columnwidth]{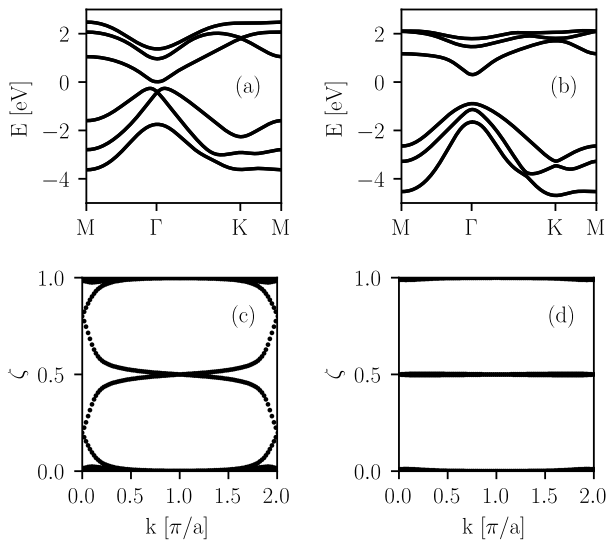}}
	\caption{Energy band structures of (a) Bi and (b) Sb bilayers. (c) and (d) are corresponding entanglement spectra, with an evidence of non-trivial topology of Bi seen as a spectral flow, not exhibited for Sb.}
	\label{fig:inf}  
\end{figure}

In Fig. \ref{fig:inf}(a) and (b) energy band structures of Bi and Sb infinite bilayers are shown. Both materials have well-defined band energy gaps around $\Gamma$ point. Topological properties of the band structures are determined by calculating the $\mathbb{Z}_2$ invariant for inversion-symmetric systems, according to the method from \cite{FuKane:Z2}. Non-trivial topology of Bi is confirm as well by observation of the spectral flow in the entanglement spectrum in Fig. \ref{fig:inf}(c). This feature is not manifested in the case of Sb, which indicates its trivial nature in Fig. \ref{fig:inf}(d).

\subsection{Bi and Sb in a ribbon geometry}
We consider a system in a ribbon geometry with periodic boundary conditions in $x$ direction, where $N_y$ denotes a width of the strip. Calculations are performed for the systems with $N_{at} = 48$ atoms. We divide the system into two parts with a cut parallel to the physical edges as depicted in Fig. \ref{fig:latt_struct}(a). Considered system size ensures that two opposite edges of the system are sufficiently far that no hybridization between potential edge states is expected and edge modes would be perfectly confined within the subsystems. 

\begin{figure}
	\centerline{\includegraphics[width=\columnwidth]{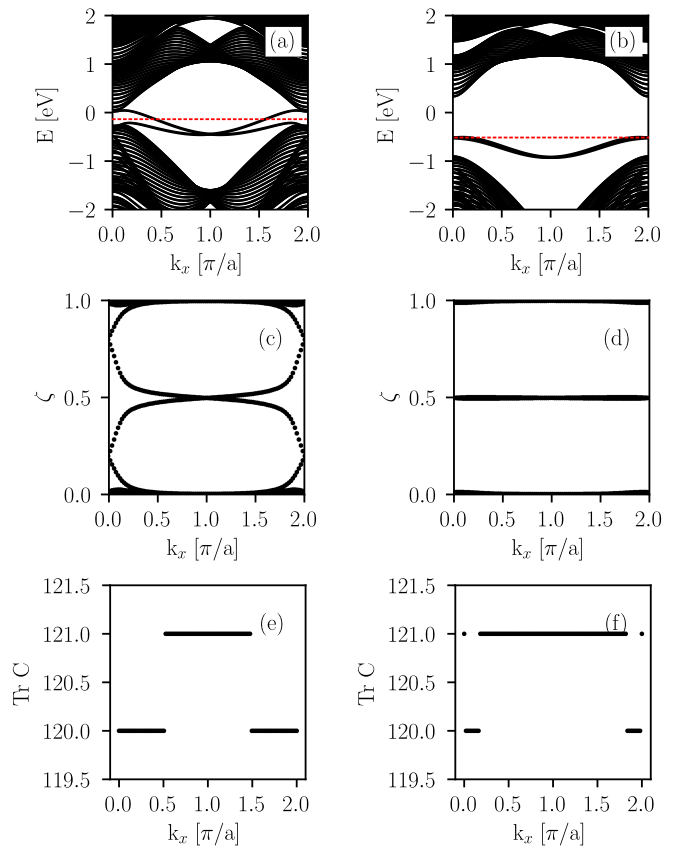}}
	\caption{(Color online) (a) and (b) Energy spectra near the Fermi level, (c) and (d) the corresponding single-particle entanglement spectra, and (e) and (f) trace indices for bismuth (left panel) and antimony (right panel) ribbons, respectively. The Fermi level is marked by red dotted lines.}
	\label{fig:rib_pan}  
\end{figure}

Fig. \ref{fig:rib_pan} compares energy spectra in the vicinity of the Fermi level ((a) and (b)), single-particle ES ((c) and (d)) and trace indices ((e) and (f)) of Bi and Sb zigzag ribbons, respectively. We look at the features characteristic of non-trivial and trivial phases. Edge states spectrally connecting the conduction and valence bands are observed in Bi bilayer, Fig. \ref{fig:rib_pan}(a), which is typical of TI regime. It is in contrast to almost flat-band edge states in Sb exhibited in the middle of the energy band gap and presented in Fig. \ref{fig:rib_pan}(b). Entanglement spectra for a zigzag ribbon do not differ in comparision to the infinite case in Fig. \ref{fig:inf}(c) and (d). We highlight that Bi armchair ribbons posses an extra pair of edge states in the energy gap, which lead to spurious modes in the single-particle ES, but the subsequent conclusions would remain the same for both edge terminations. Topological properties can be also determined by counting the number of trace discontinuities in half the BZ $mod$ $2$, which is related to $\mathbb{Z}_2$ invariant  \cite{Alex:CM}. In Fig. \ref{fig:rib_pan}(e) there is only one jump discontinuity by 1 in $k_x \in \left[ 0, \pi \right]$, which leads to nontrivial topological invariant $\mathbb{Z}_2 = 1$. On the contrary, two discontinuities in half the BZ are noticed in Fig. \ref{fig:rib_pan}(f), hence $\mathbb{Z}_2 = 0$.
 
\section{\label{sec:TPT}Topological phase transitions}
\subsection{Composition-induced phase transition in Bi$_{1-x}$Sb$_x$}
\begin{figure}
	\centerline{\includegraphics[width=0.85\columnwidth]{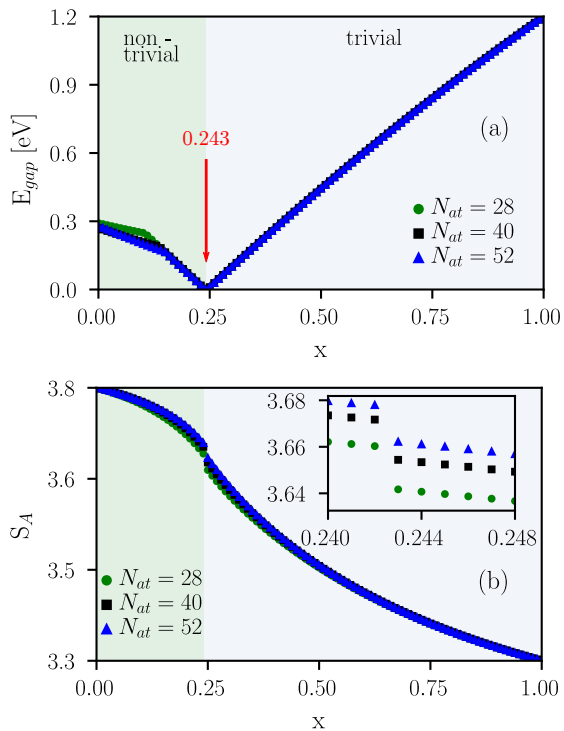}}
	\caption{(Color online) (a) Band gap evolution at the $k = 0$ point and (b) entanglement entropy as a function of antimony composition in Bi$_{1-x}$Sb$_{x}$. Red arrow in (a) indicates the band gap closing point. Different background colors refer to non-trivial/trivial regime. The inset in (b) zooms into discontinuity of $S_A$.}
	\label{fig:ent_x}  
\end{figure}

We investigate the transition to a trivial phase as the composition of antimony in Bi$_{1-x}$Sb$_{x}$ alloy increases. We consider a system in a torus geometry by applying periodic boundary conditions to the ribbon presented in Fig. \ref{fig:latt_struct}(a) also along $y$ direction. Dividing the system into two spatial parts introduces two boundaries and results in a spectral symmetry as all the single-particle ES eigenvalues come in pairs \cite{Sondhi:univ}. Three different size systems are examined: $N_y = 7, \, 10, \, 13$, corresponding to $N_{at} = 28, \, 48, \, 52$ atoms, respectively. Using virtual crystal approximation, we effectively change the values of hopping integrals between all lattice sites $t_x = (1-x) \cdot t_{\textnormal{Bi}} + x\cdot t_{\textnormal{Sb}}$, where $t_{Bi/Sb}$ are parameters from Table \ref{tab:TB}. Firstly, we look at the dependence of the energy band gap at $k = 0$ on the alloy composition, shown in Fig. \ref{fig:ent_x}(a). The energy gap of pure Bi ($x=0$) is $ E_{gap} = 0.3$ eV. Finite-size effects are noticed in the range of small amounts of antimony as a kink in the energy gap evolution around $x\sim 0.12$ for a system with $N_{at} = 28$. The energy gap decreases with an increase of Sb concentration $x$ and closes at $x = 0.243$ regardless of the system size. It reopens again as the system transits from TI to a trivial phase and increases linearly up to $E_{gap} = 1.2$ eV for pure Sb. The band gap closing point corresponds to topological phase transition and is in an agreement with calculations done for infinite 2D crystal. We note that single-particle ES for different Sb composition $x$ before and after phase transition (not shown here) do not differ qualitatively in comparison to pure Bi and Sb in Fig. \ref{fig:inf}(c) and (d). 

Fig. \ref{fig:ent_x}(b) presents dependence of the entanglement entropy of subsystem $A$, $S_A$, of the antimony composition in Bi$_{1-x}$Sb$_{x}$ alloy. Pure Bi is characterized by the largest value of the entanglement entropy, which decreases monotonically with $x$. A discontinuity at $x=0.243$ is observed for all system sizes and coincides with the energy band gap closing point seen in Fig. \ref{fig:ent_x}(a). After topological phase transition, the entanglement entropy $S_A$ still decreases to its minimal value for pure Sb. 

\subsection{Electric field-driven topological phase transition in Bi bilayer}
We apply an external electric field perpendicular to a pure Bi bilayer and observe whether distinct features are observed in the entanglement entropy and single-particle ES. An increase in $E_{Field}$ leads to the energetic separation between two sublattices and acts like a staggered potential in the Kane-Mele model \cite{KaneMele:QSHE2}. 

\begin{figure}[H]
	\centerline{\includegraphics[width=0.85\columnwidth]{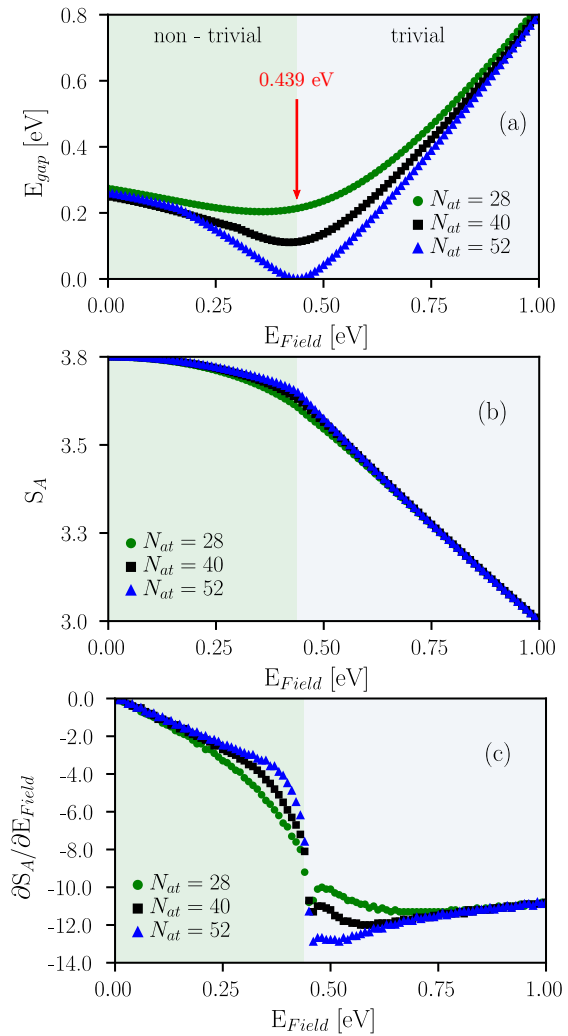}}
	\caption{(Color online) (a) Band gap evolution at the $k = 0$ as a function of electric field $E_{Field}$. For sufficiently wide system, the band gap closes at $E_{Field} = 0.439$ eV, which is indicated by a red arrow. (b) Entanglement entropy as a function of applied electric field and (c) derivative of $S_A$ with respect to $E_{Field}$. The derivative of the entanglement entropy becomes discontinuous at $E_{Field} = 0.439$ eV.}
	\label{fig:ent_Ef}  
\end{figure}

In Fig. \ref{fig:ent_Ef}(a), the energy band gap as a function of electric field $E_{Field}$ is plotted. The energy band gap does not close completely due to the finite size effects even for a very wide torus with $N_{at} = 52$ atoms, but entanglement measures are able to detect when topological phase transition should occur. In the thermodynamic limit, the band gap closes at $E_{Field} = 0.439$ eV.

Fig. \ref{fig:ent_Ef}(b) presents the entanglement entropy and (c) its first derivative with respect to $E_{Field}$. Entanglement entropy is a continuous function of $E_{Field}$, while $\partial S_A / \partial E_{Field}$ is not. An inflection point in the entanglement entropy can be seen and corresponds to the band gap closure for the infinite system. The point is more clear as we increase the system size, as sharpness of the discontinuity in the derivative is strongly size-dependent. $S_A$ decreases rapidly with the electric field after the phase transition. In the large $E_{Field}$ limit the entanglement entropy saturates to almost zero value, when sublattices can be regarded as two separate systems with no correlations between them.

\begin{figure}[H]
	\centerline{\includegraphics[width=0.95\columnwidth]{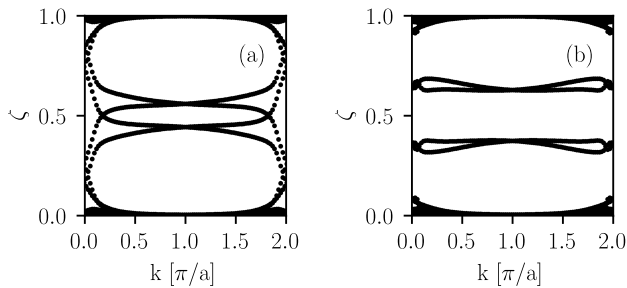}}
	\caption{(Color online) Single-particle ES (a) before, for $E_{Field}= 0.24$ eV, and (b) after, for $E_{Field}= 0.54$ eV, topological phase transition.}
	\label{fig:CM_Ef}  
\end{figure}

We look also at single-particle ES for different values of the electric field, shown in Fig. \ref{fig:CM_Ef}(a) before (for $E_{Field}= 0.24$ eV) and (b) after (for$E_{Field}= 0.54$ eV) phase transition. It differs significantly comparing to pure Bi and Sb. For small values of $E_{Field}$, initially two-fold degenerate spectrum splits into two sets of branches as illustrated in Fig. \ref{fig:CM_Ef}(a). After the phase transition to a trivial phase, the spectral flow is no longer exhibited, Fig. \ref{fig:CM_Ef}(b). Furthermore, there are no more mid-gap states at $\zeta = 0.5$, which indicates breaking of the inversion symmetry. For $E_{Field} \rightarrow \infty $ states are fully localized on the sublattices and the single-particle ES consist only of 0's and 1's, which corresponds to the non-interacting atomic limit.

\subsection{Strain-induced topological phase transition in Sb bilayer}
We consider a strain-induced phase transition in Sb bilayer. A strain is modeled by scaling hopping integrals due to change of bond lengths and angles. Following Harrison \cite{harrison}, the value of hopping parameter $V_{\alpha \beta}$ is modified as $V_{\alpha \beta } = V_{\alpha \beta}^0 \cdot \left( d / d_0 \right)^{-n}$, where $V_{\alpha \beta}^0$ corresponds to values for the unstrained case from Table I, while $d$ and $d_0$ are new and unmodified bond lengths, respectively. Here, we investigate $n = 8$ in order to enhance the effect of $V_{ij}$ strength modification. 

\begin{figure}
	\centerline{\includegraphics[width=0.85\columnwidth]{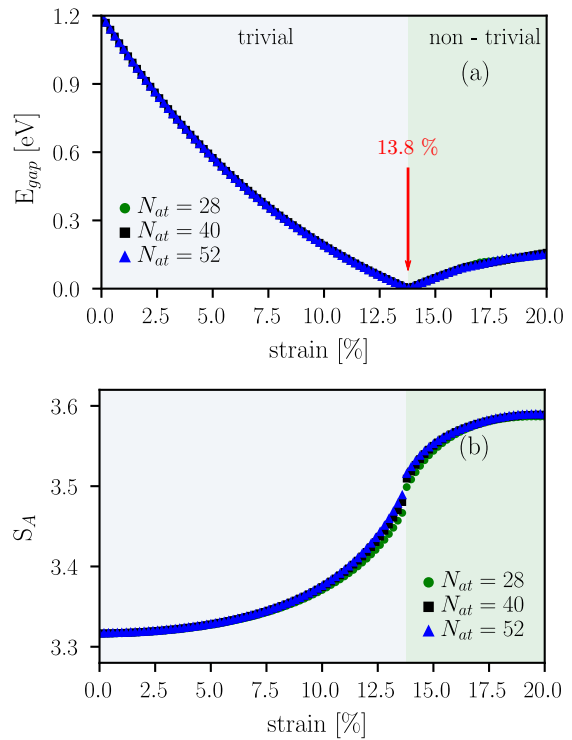}}
	\caption{(Color online) (a) Band gap evolution at $k = 0$ with respect to strain. The band gap closes at $13.8 \%$. (b) Entanglement entropy as a function of strain.}
	\label{fig:ent_s}  
\end{figure}

In Fig. \ref{fig:ent_s}(a), the band gap evolution at $k = 0$ as a function of strain is presented. The band gap decreases with a strain strength and closes at the critical value $13.8 \%$ of strain. Originally trivial antimony bilayer undergoes the topological phase transition with the bond lengthening. Since strain affects all hopping integrals between lattice sites, the inversion symmetry is preserved. Similar to the composition-induced phase transition, a small discontinuity in the entanglement entropy in Fig. \ref{fig:ent_s}(b) coincides with the band gap closing point. Also, the structure of single-particle ES before and after phase transition is almost the same as those presented in Figs. \ref{fig:inf}(c) and (d) (not shown here).

\section{\label{sec:Concl}Summary and discussion}
To conclude, we have examined two-dimensional Bi$_{1-x}$Sb$_x$ bilayers. We have confirmed their structural stability for different composition $x$ by calculations of phonon dispersion within DFT framework. We have shown that Bi$_{0.28}$Sb$_{0.72}$ and Bi$_{0.72}$Sb$_{0.28}$ thin films are structurally stable, in addition to pure bismuth and antimony bilayers, as well as Bi$_{0.5}$Sb$_{0.5}$, already discussed in the literature. Next, we have used entanglement entropy and entanglement spectrum to determine their topological properties and analyze topological phase transition. It has been shown that single-particle entanglement measures can provide supplemental information on topological properties of a system compared to electronic structure studies. We have shown that entanglement spectrum of topologically nontrivial Bi reveals a spectral flow, which is not present in trivial Sb. By considering a system in a ribbon geometry, we have shown also that topological properties of the system can be determined by looking at the trace index in half the BZ, which exhibit a single jump (odd) in a case of TI or two jumps (even) in a trivial case.

We have analyzed also phase transitions between topologically non-trivial/trivial phase driven by antimony composition $x$ in Bi$_{1-x}$Sb$_{x}$, due to applied electric field in pure Bi, and strain in pure Sb. A phase transition resulting in a composition change occurs at $x = 24.3 \, \%$, which differs from a TI regime in three-dimensional bulk Bi$_{1-x}$Sb$_{x}$, reported for $x$ ranging from $7 \%$ to $22 \%$ \cite{FuKane:Z2}. We point out that in 3D the band inversion process occurs at the $L$ point in the BZ, which maps onto $\bar{M}$ point on the surface, while the band gap is observed at the $\Gamma$ point in 2D. Composition- and strain-induced phase transitions reveal a finite discontinuity in the entanglement entropy. In the case of electric field, a phase transition seems to have a different character as entanglement entropy remains a continuous function of the electric field strength, while its first derivative is discontinuous. We relate this difference to breaking of inversion symmetry in the last case. We do not qualify whether we observe the first and second order phase transitions, respectively, as this requires more careful analysis, including behavior of correlation length at a phase transition point and we leave it for a further work.

\begin{acknowledgments} 
The authors acknowledge partial financial support from National Science Center (NCN), Poland, grant Sonata No. 2013/11/D/ST3/02703. 
\end{acknowledgments}

\bibliography{ent_bisb2.bib}

\begin{thebibliography}{63}%
\makeatletter
\providecommand \@ifxundefined [1]{%
 \@ifx{#1\undefined}
}%
\providecommand \@ifnum [1]{%
 \ifnum #1\expandafter \@firstoftwo
 \else \expandafter \@secondoftwo
 \fi
}%
\providecommand \@ifx [1]{%
 \ifx #1\expandafter \@firstoftwo
 \else \expandafter \@secondoftwo
 \fi
}%
\providecommand \natexlab [1]{#1}%
\providecommand \enquote  [1]{``#1''}%
\providecommand \bibnamefont  [1]{#1}%
\providecommand \bibfnamefont [1]{#1}%
\providecommand \citenamefont [1]{#1}%
\providecommand \href@noop [0]{\@secondoftwo}%
\providecommand \href [0]{\begingroup \@sanitize@url \@href}%
\providecommand \@href[1]{\@@startlink{#1}\@@href}%
\providecommand \@@href[1]{\endgroup#1\@@endlink}%
\providecommand \@sanitize@url [0]{\catcode `\\12\catcode `\$12\catcode
  `\&12\catcode `\#12\catcode `\^12\catcode `\_12\catcode `\%12\relax}%
\providecommand \@@startlink[1]{}%
\providecommand \@@endlink[0]{}%
\providecommand \url  [0]{\begingroup\@sanitize@url \@url }%
\providecommand \@url [1]{\endgroup\@href {#1}{\urlprefix }}%
\providecommand \urlprefix  [0]{URL }%
\providecommand \Eprint [0]{\href }%
\providecommand \doibase [0]{http://dx.doi.org/}%
\providecommand \selectlanguage [0]{\@gobble}%
\providecommand \bibinfo  [0]{\@secondoftwo}%
\providecommand \bibfield  [0]{\@secondoftwo}%
\providecommand \translation [1]{[#1]}%
\providecommand \BibitemOpen [0]{}%
\providecommand \bibitemStop [0]{}%
\providecommand \bibitemNoStop [0]{.\EOS\space}%
\providecommand \EOS [0]{\spacefactor3000\relax}%
\providecommand \BibitemShut  [1]{\csname bibitem#1\endcsname}%
\let\auto@bib@innerbib\@empty
\bibitem [{\citenamefont {Hasan}\ and\ \citenamefont {Kane}(2010)}]{Hasan:rev}%
  \BibitemOpen
  \bibfield  {author} {\bibinfo {author} {\bibfnamefont {M.~Z.}\ \bibnamefont
  {Hasan}}\ and\ \bibinfo {author} {\bibfnamefont {C.~L.}\ \bibnamefont
  {Kane}},\ }\href {\doibase 10.1103/RevModPhys.82.3045} {\bibfield  {journal}
  {\bibinfo  {journal} {Rev. Mod. Phys.}\ }\textbf {\bibinfo {volume} {82}},\
  \bibinfo {pages} {3045} (\bibinfo {year} {2010})}\BibitemShut {NoStop}%
\bibitem [{\citenamefont {Qi}\ and\ \citenamefont {Zhang}(2011)}]{Zhang:rev}%
  \BibitemOpen
  \bibfield  {author} {\bibinfo {author} {\bibfnamefont {X.-L.}\ \bibnamefont
  {Qi}}\ and\ \bibinfo {author} {\bibfnamefont {S.-C.}\ \bibnamefont {Zhang}},\
  }\href {\doibase 10.1103/RevModPhys.83.1057} {\bibfield  {journal} {\bibinfo
  {journal} {Rev. Mod. Phys.}\ }\textbf {\bibinfo {volume} {83}},\ \bibinfo
  {pages} {1057} (\bibinfo {year} {2011})}\BibitemShut {NoStop}%
\bibitem [{\citenamefont {Nayak}\ \emph {et~al.}(2008)\citenamefont {Nayak},
  \citenamefont {Simon}, \citenamefont {Stern}, \citenamefont {Freedman},\ and\
  \citenamefont {Das~Sarma}}]{QComp}%
  \BibitemOpen
  \bibfield  {author} {\bibinfo {author} {\bibfnamefont {C.}~\bibnamefont
  {Nayak}}, \bibinfo {author} {\bibfnamefont {S.~H.}\ \bibnamefont {Simon}},
  \bibinfo {author} {\bibfnamefont {A.}~\bibnamefont {Stern}}, \bibinfo
  {author} {\bibfnamefont {M.}~\bibnamefont {Freedman}}, \ and\ \bibinfo
  {author} {\bibfnamefont {S.}~\bibnamefont {Das~Sarma}},\ }\href {\doibase
  10.1103/RevModPhys.80.1083} {\bibfield  {journal} {\bibinfo  {journal} {Rev.
  Mod. Phys.}\ }\textbf {\bibinfo {volume} {80}},\ \bibinfo {pages} {1083}
  (\bibinfo {year} {2008})}\BibitemShut {NoStop}%
\bibitem [{\citenamefont {Kane}\ and\ \citenamefont
  {Mele}(2005{\natexlab{a}})}]{KaneMele:QSHE}%
  \BibitemOpen
  \bibfield  {author} {\bibinfo {author} {\bibfnamefont {C.~L.}\ \bibnamefont
  {Kane}}\ and\ \bibinfo {author} {\bibfnamefont {E.~J.}\ \bibnamefont
  {Mele}},\ }\href {\doibase 10.1103/PhysRevLett.95.146802} {\bibfield
  {journal} {\bibinfo  {journal} {Phys. Rev. Lett.}\ }\textbf {\bibinfo
  {volume} {95}},\ \bibinfo {pages} {146802} (\bibinfo {year}
  {2005}{\natexlab{a}})}\BibitemShut {NoStop}%
\bibitem [{\citenamefont {Schnyder}\ \emph {et~al.}(2008)\citenamefont
  {Schnyder}, \citenamefont {Ryu}, \citenamefont {Furusaki},\ and\
  \citenamefont {Ludwig}}]{Schnyder:class}%
  \BibitemOpen
  \bibfield  {author} {\bibinfo {author} {\bibfnamefont {A.~P.}\ \bibnamefont
  {Schnyder}}, \bibinfo {author} {\bibfnamefont {S.}~\bibnamefont {Ryu}},
  \bibinfo {author} {\bibfnamefont {A.}~\bibnamefont {Furusaki}}, \ and\
  \bibinfo {author} {\bibfnamefont {A.~W.~W.}\ \bibnamefont {Ludwig}},\ }\href
  {\doibase 10.1103/PhysRevB.78.195125} {\bibfield  {journal} {\bibinfo
  {journal} {Phys. Rev. B}\ }\textbf {\bibinfo {volume} {78}},\ \bibinfo
  {pages} {195125} (\bibinfo {year} {2008})}\BibitemShut {NoStop}%
\bibitem [{\citenamefont {Fu}\ and\ \citenamefont {Kane}(2007)}]{FuKane:Z2}%
  \BibitemOpen
  \bibfield  {author} {\bibinfo {author} {\bibfnamefont {L.}~\bibnamefont
  {Fu}}\ and\ \bibinfo {author} {\bibfnamefont {C.~L.}\ \bibnamefont {Kane}},\
  }\href {\doibase 10.1103/PhysRevB.76.045302} {\bibfield  {journal} {\bibinfo
  {journal} {Phys. Rev. B}\ }\textbf {\bibinfo {volume} {76}},\ \bibinfo
  {pages} {045302} (\bibinfo {year} {2007})}\BibitemShut {NoStop}%
\bibitem [{\citenamefont {{Potasz}}\ and\ \citenamefont
  {{Fern{\'a}ndez-Rossier}}(2015)}]{potasz}%
  \BibitemOpen
  \bibfield  {author} {\bibinfo {author} {\bibfnamefont {P.}~\bibnamefont
  {{Potasz}}}\ and\ \bibinfo {author} {\bibfnamefont {J.}~\bibnamefont
  {{Fern{\'a}ndez-Rossier}}},\ }\href {\doibase 10.1021/acs.nanolett.5b01805}
  {\bibfield  {journal} {\bibinfo  {journal} {Nano Letters}\ }\textbf {\bibinfo
  {volume} {15}},\ \bibinfo {pages} {5799} (\bibinfo {year}
  {2015})}\BibitemShut {NoStop}%
\bibitem [{\citenamefont {Bernevig}\ and\ \citenamefont
  {Zhang}(2006)}]{QSHE:Bernevig}%
  \BibitemOpen
  \bibfield  {author} {\bibinfo {author} {\bibfnamefont {B.~A.}\ \bibnamefont
  {Bernevig}}\ and\ \bibinfo {author} {\bibfnamefont {S.-C.}\ \bibnamefont
  {Zhang}},\ }\href {\doibase 10.1103/PhysRevLett.96.106802} {\bibfield
  {journal} {\bibinfo  {journal} {Phys. Rev. Lett.}\ }\textbf {\bibinfo
  {volume} {96}},\ \bibinfo {pages} {106802} (\bibinfo {year}
  {2006})}\BibitemShut {NoStop}%
\bibitem [{\citenamefont {Bernevig}\ \emph {et~al.}(2006)\citenamefont
  {Bernevig}, \citenamefont {Hughes},\ and\ \citenamefont {Zhang}}]{BHZ}%
  \BibitemOpen
  \bibfield  {author} {\bibinfo {author} {\bibfnamefont {B.~A.}\ \bibnamefont
  {Bernevig}}, \bibinfo {author} {\bibfnamefont {T.~L.}\ \bibnamefont
  {Hughes}}, \ and\ \bibinfo {author} {\bibfnamefont {S.-C.}\ \bibnamefont
  {Zhang}},\ }\href {\doibase 10.1126/science.1133734} {\bibfield  {journal}
  {\bibinfo  {journal} {Science}\ }\textbf {\bibinfo {volume} {314}},\ \bibinfo
  {pages} {1757} (\bibinfo {year} {2006})}\BibitemShut {NoStop}%
\bibitem [{\citenamefont {Liu}\ \emph {et~al.}(2008)\citenamefont {Liu},
  \citenamefont {Hughes}, \citenamefont {Qi}, \citenamefont {Wang},\ and\
  \citenamefont {Zhang}}]{QSHE:Zhang}%
  \BibitemOpen
  \bibfield  {author} {\bibinfo {author} {\bibfnamefont {C.}~\bibnamefont
  {Liu}}, \bibinfo {author} {\bibfnamefont {T.~L.}\ \bibnamefont {Hughes}},
  \bibinfo {author} {\bibfnamefont {X.-L.}\ \bibnamefont {Qi}}, \bibinfo
  {author} {\bibfnamefont {K.}~\bibnamefont {Wang}}, \ and\ \bibinfo {author}
  {\bibfnamefont {S.-C.}\ \bibnamefont {Zhang}},\ }\href {\doibase
  10.1103/PhysRevLett.100.236601} {\bibfield  {journal} {\bibinfo  {journal}
  {Phys. Rev. Lett.}\ }\textbf {\bibinfo {volume} {100}},\ \bibinfo {pages}
  {236601} (\bibinfo {year} {2008})}\BibitemShut {NoStop}%
\bibitem [{\citenamefont {Ren}\ \emph {et~al.}(2016)\citenamefont {Ren},
  \citenamefont {Qiao},\ and\ \citenamefont {Niu}}]{ren2016topological}%
  \BibitemOpen
  \bibfield  {author} {\bibinfo {author} {\bibfnamefont {Y.}~\bibnamefont
  {Ren}}, \bibinfo {author} {\bibfnamefont {Z.}~\bibnamefont {Qiao}}, \ and\
  \bibinfo {author} {\bibfnamefont {Q.}~\bibnamefont {Niu}},\ }\href@noop {}
  {\bibfield  {journal} {\bibinfo  {journal} {Reports on Progress in Physics}\
  }\textbf {\bibinfo {volume} {79}},\ \bibinfo {pages} {066501} (\bibinfo
  {year} {2016})}\BibitemShut {NoStop}%
\bibitem [{\citenamefont {Safaei}\ \emph {et~al.}(2015)\citenamefont {Safaei},
  \citenamefont {Galicka}, \citenamefont {Kacman},\ and\ \citenamefont
  {Buczko}}]{TCI}%
  \BibitemOpen
  \bibfield  {author} {\bibinfo {author} {\bibfnamefont {S.}~\bibnamefont
  {Safaei}}, \bibinfo {author} {\bibfnamefont {M.}~\bibnamefont {Galicka}},
  \bibinfo {author} {\bibfnamefont {P.}~\bibnamefont {Kacman}}, \ and\ \bibinfo
  {author} {\bibfnamefont {R.}~\bibnamefont {Buczko}},\ }\href
  {http://stacks.iop.org/1367-2630/17/i=6/a=063041} {\bibfield  {journal}
  {\bibinfo  {journal} {New Journal of Physics}\ }\textbf {\bibinfo {volume}
  {17}},\ \bibinfo {pages} {063041} (\bibinfo {year} {2015})}\BibitemShut
  {NoStop}%
\bibitem [{\citenamefont {Zhou}\ and\ \citenamefont {Jena}(2017)}]{TCI2}%
  \BibitemOpen
  \bibfield  {author} {\bibinfo {author} {\bibfnamefont {J.}~\bibnamefont
  {Zhou}}\ and\ \bibinfo {author} {\bibfnamefont {P.}~\bibnamefont {Jena}},\
  }\href {\doibase 10.1103/PhysRevB.95.081102} {\bibfield  {journal} {\bibinfo
  {journal} {Phys. Rev. B}\ }\textbf {\bibinfo {volume} {95}},\ \bibinfo
  {pages} {081102} (\bibinfo {year} {2017})}\BibitemShut {NoStop}%
\bibitem [{\citenamefont {Murakami}(2006)}]{Murakami:BiQSH}%
  \BibitemOpen
  \bibfield  {author} {\bibinfo {author} {\bibfnamefont {S.}~\bibnamefont
  {Murakami}},\ }\href {\doibase 10.1103/PhysRevLett.97.236805} {\bibfield
  {journal} {\bibinfo  {journal} {Phys. Rev. Lett.}\ }\textbf {\bibinfo
  {volume} {97}},\ \bibinfo {pages} {236805} (\bibinfo {year}
  {2006})}\BibitemShut {NoStop}%
\bibitem [{\citenamefont {Liu}\ \emph {et~al.}(2011)\citenamefont {Liu},
  \citenamefont {Liu}, \citenamefont {Wu}, \citenamefont {Duan}, \citenamefont
  {Liu},\ and\ \citenamefont {Wu}}]{stable:bi111}%
  \BibitemOpen
  \bibfield  {author} {\bibinfo {author} {\bibfnamefont {Z.}~\bibnamefont
  {Liu}}, \bibinfo {author} {\bibfnamefont {C.-X.}\ \bibnamefont {Liu}},
  \bibinfo {author} {\bibfnamefont {Y.-S.}\ \bibnamefont {Wu}}, \bibinfo
  {author} {\bibfnamefont {W.-H.}\ \bibnamefont {Duan}}, \bibinfo {author}
  {\bibfnamefont {F.}~\bibnamefont {Liu}}, \ and\ \bibinfo {author}
  {\bibfnamefont {J.}~\bibnamefont {Wu}},\ }\href {\doibase
  10.1103/PhysRevLett.107.136805} {\bibfield  {journal} {\bibinfo  {journal}
  {Phys. Rev. Lett.}\ }\textbf {\bibinfo {volume} {107}},\ \bibinfo {pages}
  {136805} (\bibinfo {year} {2011})}\BibitemShut {NoStop}%
\bibitem [{\citenamefont {Huang}\ \emph {et~al.}(2013)\citenamefont {Huang},
  \citenamefont {Chuang}, \citenamefont {Hsu}, \citenamefont {Liu},
  \citenamefont {Chang}, \citenamefont {Lin},\ and\ \citenamefont
  {Bansil}}]{singlebi111}%
  \BibitemOpen
  \bibfield  {author} {\bibinfo {author} {\bibfnamefont {Z.-Q.}\ \bibnamefont
  {Huang}}, \bibinfo {author} {\bibfnamefont {F.-C.}\ \bibnamefont {Chuang}},
  \bibinfo {author} {\bibfnamefont {C.-H.}\ \bibnamefont {Hsu}}, \bibinfo
  {author} {\bibfnamefont {Y.-T.}\ \bibnamefont {Liu}}, \bibinfo {author}
  {\bibfnamefont {H.-R.}\ \bibnamefont {Chang}}, \bibinfo {author}
  {\bibfnamefont {H.}~\bibnamefont {Lin}}, \ and\ \bibinfo {author}
  {\bibfnamefont {A.}~\bibnamefont {Bansil}},\ }\href {\doibase
  10.1103/PhysRevB.88.165301} {\bibfield  {journal} {\bibinfo  {journal} {Phys.
  Rev. B}\ }\textbf {\bibinfo {volume} {88}},\ \bibinfo {pages} {165301}
  (\bibinfo {year} {2013})}\BibitemShut {NoStop}%
\bibitem [{\citenamefont {Wang}\ \emph {et~al.}(2013)\citenamefont {Wang},
  \citenamefont {Chen}, \citenamefont {Liu},\ and\ \citenamefont
  {Wang}}]{ElTPbi}%
  \BibitemOpen
  \bibfield  {author} {\bibinfo {author} {\bibfnamefont {D.}~\bibnamefont
  {Wang}}, \bibinfo {author} {\bibfnamefont {L.}~\bibnamefont {Chen}}, \bibinfo
  {author} {\bibfnamefont {H.}~\bibnamefont {Liu}}, \ and\ \bibinfo {author}
  {\bibfnamefont {X.}~\bibnamefont {Wang}},\ }\href {\doibase
  10.7566/JPSJ.82.094712} {\bibfield  {journal} {\bibinfo  {journal} {Journal
  of the Physical Society of Japan}\ }\textbf {\bibinfo {volume} {82}},\
  \bibinfo {pages} {094712} (\bibinfo {year} {2013})}\BibitemShut {NoStop}%
\bibitem [{\citenamefont {Drozdov}\ \emph {et~al.}(2014)\citenamefont
  {Drozdov}, \citenamefont {Alexandradinata}, \citenamefont {Jeon},
  \citenamefont {Nadj-Perge}, \citenamefont {Ji}, \citenamefont {Cava},
  \citenamefont {Bernevig},\ and\ \citenamefont {Yazdani}}]{drozdov}%
  \BibitemOpen
  \bibfield  {author} {\bibinfo {author} {\bibfnamefont {I.~K.}\ \bibnamefont
  {Drozdov}}, \bibinfo {author} {\bibfnamefont {A.}~\bibnamefont
  {Alexandradinata}}, \bibinfo {author} {\bibfnamefont {S.}~\bibnamefont
  {Jeon}}, \bibinfo {author} {\bibfnamefont {S.}~\bibnamefont {Nadj-Perge}},
  \bibinfo {author} {\bibfnamefont {H.}~\bibnamefont {Ji}}, \bibinfo {author}
  {\bibfnamefont {R.}~\bibnamefont {Cava}}, \bibinfo {author} {\bibfnamefont
  {B.~A.}\ \bibnamefont {Bernevig}}, \ and\ \bibinfo {author} {\bibfnamefont
  {A.}~\bibnamefont {Yazdani}},\ }\href@noop {} {\bibfield  {journal} {\bibinfo
   {journal} {Nature Physics}\ }\textbf {\bibinfo {volume} {10}},\ \bibinfo
  {pages} {664} (\bibinfo {year} {2014})}\BibitemShut {NoStop}%
\bibitem [{\citenamefont {Kawakami}\ \emph {et~al.}(2015)\citenamefont
  {Kawakami}, \citenamefont {Lin}, \citenamefont {Kawai}, \citenamefont
  {Arafune},\ and\ \citenamefont {Takagi}}]{kawakami2015one}%
  \BibitemOpen
  \bibfield  {author} {\bibinfo {author} {\bibfnamefont {N.}~\bibnamefont
  {Kawakami}}, \bibinfo {author} {\bibfnamefont {C.-L.}\ \bibnamefont {Lin}},
  \bibinfo {author} {\bibfnamefont {M.}~\bibnamefont {Kawai}}, \bibinfo
  {author} {\bibfnamefont {R.}~\bibnamefont {Arafune}}, \ and\ \bibinfo
  {author} {\bibfnamefont {N.}~\bibnamefont {Takagi}},\ }\href@noop {}
  {\bibfield  {journal} {\bibinfo  {journal} {Applied Physics Letters}\
  }\textbf {\bibinfo {volume} {107}},\ \bibinfo {pages} {031602} (\bibinfo
  {year} {2015})}\BibitemShut {NoStop}%
\bibitem [{\citenamefont {Yang}\ \emph {et~al.}(2012)\citenamefont {Yang},
  \citenamefont {Miao}, \citenamefont {Wang}, \citenamefont {Yao},
  \citenamefont {Zhu}, \citenamefont {Song}, \citenamefont {Wang},
  \citenamefont {Xu}, \citenamefont {Fedorov}, \citenamefont {Sun},
  \citenamefont {Zhang}, \citenamefont {Liu}, \citenamefont {Liu},
  \citenamefont {Qian}, \citenamefont {Gao},\ and\ \citenamefont
  {Jia}}]{spatialandenergy}%
  \BibitemOpen
  \bibfield  {author} {\bibinfo {author} {\bibfnamefont {F.}~\bibnamefont
  {Yang}}, \bibinfo {author} {\bibfnamefont {L.}~\bibnamefont {Miao}}, \bibinfo
  {author} {\bibfnamefont {Z.~F.}\ \bibnamefont {Wang}}, \bibinfo {author}
  {\bibfnamefont {M.-Y.}\ \bibnamefont {Yao}}, \bibinfo {author} {\bibfnamefont
  {F.}~\bibnamefont {Zhu}}, \bibinfo {author} {\bibfnamefont {Y.~R.}\
  \bibnamefont {Song}}, \bibinfo {author} {\bibfnamefont {M.-X.}\ \bibnamefont
  {Wang}}, \bibinfo {author} {\bibfnamefont {J.-P.}\ \bibnamefont {Xu}},
  \bibinfo {author} {\bibfnamefont {A.~V.}\ \bibnamefont {Fedorov}}, \bibinfo
  {author} {\bibfnamefont {Z.}~\bibnamefont {Sun}}, \bibinfo {author}
  {\bibfnamefont {G.~B.}\ \bibnamefont {Zhang}}, \bibinfo {author}
  {\bibfnamefont {C.}~\bibnamefont {Liu}}, \bibinfo {author} {\bibfnamefont
  {F.}~\bibnamefont {Liu}}, \bibinfo {author} {\bibfnamefont {D.}~\bibnamefont
  {Qian}}, \bibinfo {author} {\bibfnamefont {C.~L.}\ \bibnamefont {Gao}}, \
  and\ \bibinfo {author} {\bibfnamefont {J.-F.}\ \bibnamefont {Jia}},\ }\href
  {\doibase 10.1103/PhysRevLett.109.016801} {\bibfield  {journal} {\bibinfo
  {journal} {Phys. Rev. Lett.}\ }\textbf {\bibinfo {volume} {109}},\ \bibinfo
  {pages} {016801} (\bibinfo {year} {2012})}\BibitemShut {NoStop}%
\bibitem [{\citenamefont {Taskin}\ \emph {et~al.}(2012)\citenamefont {Taskin},
  \citenamefont {Sasaki}, \citenamefont {Segawa},\ and\ \citenamefont
  {Ando}}]{BiSe2d}%
  \BibitemOpen
  \bibfield  {author} {\bibinfo {author} {\bibfnamefont {A.~A.}\ \bibnamefont
  {Taskin}}, \bibinfo {author} {\bibfnamefont {S.}~\bibnamefont {Sasaki}},
  \bibinfo {author} {\bibfnamefont {K.}~\bibnamefont {Segawa}}, \ and\ \bibinfo
  {author} {\bibfnamefont {Y.}~\bibnamefont {Ando}},\ }\href {\doibase
  10.1103/PhysRevLett.109.066803} {\bibfield  {journal} {\bibinfo  {journal}
  {Phys. Rev. Lett.}\ }\textbf {\bibinfo {volume} {109}},\ \bibinfo {pages}
  {066803} (\bibinfo {year} {2012})}\BibitemShut {NoStop}%
\bibitem [{\citenamefont {Liu}\ \emph {et~al.}(2010)\citenamefont {Liu},
  \citenamefont {Zhang}, \citenamefont {Yan}, \citenamefont {Qi}, \citenamefont
  {Frauenheim}, \citenamefont {Dai}, \citenamefont {Fang},\ and\ \citenamefont
  {Zhang}}]{Zhang:1}%
  \BibitemOpen
  \bibfield  {author} {\bibinfo {author} {\bibfnamefont {C.-X.}\ \bibnamefont
  {Liu}}, \bibinfo {author} {\bibfnamefont {H.}~\bibnamefont {Zhang}}, \bibinfo
  {author} {\bibfnamefont {B.}~\bibnamefont {Yan}}, \bibinfo {author}
  {\bibfnamefont {X.-L.}\ \bibnamefont {Qi}}, \bibinfo {author} {\bibfnamefont
  {T.}~\bibnamefont {Frauenheim}}, \bibinfo {author} {\bibfnamefont
  {X.}~\bibnamefont {Dai}}, \bibinfo {author} {\bibfnamefont {Z.}~\bibnamefont
  {Fang}}, \ and\ \bibinfo {author} {\bibfnamefont {S.-C.}\ \bibnamefont
  {Zhang}},\ }\href {\doibase 10.1103/PhysRevB.81.041307} {\bibfield  {journal}
  {\bibinfo  {journal} {Phys. Rev. B}\ }\textbf {\bibinfo {volume} {81}},\
  \bibinfo {pages} {041307} (\bibinfo {year} {2010})}\BibitemShut {NoStop}%
\bibitem [{\citenamefont {Hirahara}\ \emph {et~al.}(2011)\citenamefont
  {Hirahara}, \citenamefont {Bihlmayer}, \citenamefont {Sakamoto},
  \citenamefont {Yamada}, \citenamefont {Miyazaki}, \citenamefont {Kimura},
  \citenamefont {Bl\"ugel},\ and\ \citenamefont {Hasegawa}}]{BiTe3inter}%
  \BibitemOpen
  \bibfield  {author} {\bibinfo {author} {\bibfnamefont {T.}~\bibnamefont
  {Hirahara}}, \bibinfo {author} {\bibfnamefont {G.}~\bibnamefont {Bihlmayer}},
  \bibinfo {author} {\bibfnamefont {Y.}~\bibnamefont {Sakamoto}}, \bibinfo
  {author} {\bibfnamefont {M.}~\bibnamefont {Yamada}}, \bibinfo {author}
  {\bibfnamefont {H.}~\bibnamefont {Miyazaki}}, \bibinfo {author}
  {\bibfnamefont {S.-i.}\ \bibnamefont {Kimura}}, \bibinfo {author}
  {\bibfnamefont {S.}~\bibnamefont {Bl\"ugel}}, \ and\ \bibinfo {author}
  {\bibfnamefont {S.}~\bibnamefont {Hasegawa}},\ }\href {\doibase
  10.1103/PhysRevLett.107.166801} {\bibfield  {journal} {\bibinfo  {journal}
  {Phys. Rev. Lett.}\ }\textbf {\bibinfo {volume} {107}},\ \bibinfo {pages}
  {166801} (\bibinfo {year} {2011})}\BibitemShut {NoStop}%
\bibitem [{\citenamefont {Bieniek}\ \emph {et~al.}(2017)\citenamefont
  {Bieniek}, \citenamefont {Wo\'{z}niak},\ and\ \citenamefont
  {Potasz}}]{stability:bi111}%
  \BibitemOpen
  \bibfield  {author} {\bibinfo {author} {\bibfnamefont {M.}~\bibnamefont
  {Bieniek}}, \bibinfo {author} {\bibfnamefont {T.}~\bibnamefont
  {Wo\'{z}niak}}, \ and\ \bibinfo {author} {\bibfnamefont {P.}~\bibnamefont
  {Potasz}},\ }\href {http://stacks.iop.org/0953-8984/29/i=15/a=155501}
  {\bibfield  {journal} {\bibinfo  {journal} {Journal of Physics: Condensed
  Matter}\ }\textbf {\bibinfo {volume} {29}},\ \bibinfo {pages} {155501}
  (\bibinfo {year} {2017})}\BibitemShut {NoStop}%
\bibitem [{\citenamefont {Chen}\ \emph {et~al.}(2013)\citenamefont {Chen},
  \citenamefont {Wang},\ and\ \citenamefont {Liu}}]{Bi:robust}%
  \BibitemOpen
  \bibfield  {author} {\bibinfo {author} {\bibfnamefont {L.}~\bibnamefont
  {Chen}}, \bibinfo {author} {\bibfnamefont {Z.~F.}\ \bibnamefont {Wang}}, \
  and\ \bibinfo {author} {\bibfnamefont {F.}~\bibnamefont {Liu}},\ }\href
  {\doibase 10.1103/PhysRevB.87.235420} {\bibfield  {journal} {\bibinfo
  {journal} {Phys. Rev. B}\ }\textbf {\bibinfo {volume} {87}},\ \bibinfo
  {pages} {235420} (\bibinfo {year} {2013})}\BibitemShut {NoStop}%
\bibitem [{\citenamefont {Li}\ \emph {et~al.}(2014)\citenamefont {Li},
  \citenamefont {Liu}, \citenamefont {Jiang}, \citenamefont {Wang},\ and\
  \citenamefont {Feng}}]{edgeen}%
  \BibitemOpen
  \bibfield  {author} {\bibinfo {author} {\bibfnamefont {X.}~\bibnamefont
  {Li}}, \bibinfo {author} {\bibfnamefont {H.}~\bibnamefont {Liu}}, \bibinfo
  {author} {\bibfnamefont {H.}~\bibnamefont {Jiang}}, \bibinfo {author}
  {\bibfnamefont {F.}~\bibnamefont {Wang}}, \ and\ \bibinfo {author}
  {\bibfnamefont {J.}~\bibnamefont {Feng}},\ }\href {\doibase
  10.1103/PhysRevB.90.165412} {\bibfield  {journal} {\bibinfo  {journal} {Phys.
  Rev. B}\ }\textbf {\bibinfo {volume} {90}},\ \bibinfo {pages} {165412}
  (\bibinfo {year} {2014})}\BibitemShut {NoStop}%
\bibitem [{\citenamefont {Koroteev}\ \emph {et~al.}(2008)\citenamefont
  {Koroteev}, \citenamefont {Bihlmayer}, \citenamefont {Chulkov},\ and\
  \citenamefont {Bl\"ugel}}]{koroteev}%
  \BibitemOpen
  \bibfield  {author} {\bibinfo {author} {\bibfnamefont {Y.~M.}\ \bibnamefont
  {Koroteev}}, \bibinfo {author} {\bibfnamefont {G.}~\bibnamefont {Bihlmayer}},
  \bibinfo {author} {\bibfnamefont {E.~V.}\ \bibnamefont {Chulkov}}, \ and\
  \bibinfo {author} {\bibfnamefont {S.}~\bibnamefont {Bl\"ugel}},\ }\href
  {\doibase 10.1103/PhysRevB.77.045428} {\bibfield  {journal} {\bibinfo
  {journal} {Phys. Rev. B}\ }\textbf {\bibinfo {volume} {77}},\ \bibinfo
  {pages} {045428} (\bibinfo {year} {2008})}\BibitemShut {NoStop}%
\bibitem [{\citenamefont {Pan}\ and\ \citenamefont
  {Wang}(2015)}]{pan2015realization}%
  \BibitemOpen
  \bibfield  {author} {\bibinfo {author} {\bibfnamefont {H.}~\bibnamefont
  {Pan}}\ and\ \bibinfo {author} {\bibfnamefont {X.-S.}\ \bibnamefont {Wang}},\
  }\href@noop {} {\bibfield  {journal} {\bibinfo  {journal} {Nanoscale research
  letters}\ }\textbf {\bibinfo {volume} {10}},\ \bibinfo {pages} {334}
  (\bibinfo {year} {2015})}\BibitemShut {NoStop}%
\bibitem [{\citenamefont {Bian}\ \emph {et~al.}(2016)\citenamefont {Bian},
  \citenamefont {Wang}, \citenamefont {Wang}, \citenamefont {Xu}, \citenamefont
  {Xu}, \citenamefont {Miller}, \citenamefont {Hasan}, \citenamefont {Liu},\
  and\ \citenamefont {Chiang}}]{bian2016engineering}%
  \BibitemOpen
  \bibfield  {author} {\bibinfo {author} {\bibfnamefont {G.}~\bibnamefont
  {Bian}}, \bibinfo {author} {\bibfnamefont {Z.}~\bibnamefont {Wang}}, \bibinfo
  {author} {\bibfnamefont {X.-X.}\ \bibnamefont {Wang}}, \bibinfo {author}
  {\bibfnamefont {C.}~\bibnamefont {Xu}}, \bibinfo {author} {\bibfnamefont
  {S.}~\bibnamefont {Xu}}, \bibinfo {author} {\bibfnamefont {T.}~\bibnamefont
  {Miller}}, \bibinfo {author} {\bibfnamefont {M.~Z.}\ \bibnamefont {Hasan}},
  \bibinfo {author} {\bibfnamefont {F.}~\bibnamefont {Liu}}, \ and\ \bibinfo
  {author} {\bibfnamefont {T.-C.}\ \bibnamefont {Chiang}},\ }\href@noop {}
  {\bibfield  {journal} {\bibinfo  {journal} {ACS nano}\ }\textbf {\bibinfo
  {volume} {10}},\ \bibinfo {pages} {3859} (\bibinfo {year}
  {2016})}\BibitemShut {NoStop}%
\bibitem [{\citenamefont {Cantele}\ and\ \citenamefont
  {Ninno}(2017)}]{PhysRevMaterials.1.014002}%
  \BibitemOpen
  \bibfield  {author} {\bibinfo {author} {\bibfnamefont {G.}~\bibnamefont
  {Cantele}}\ and\ \bibinfo {author} {\bibfnamefont {D.}~\bibnamefont
  {Ninno}},\ }\href {\doibase 10.1103/PhysRevMaterials.1.014002} {\bibfield
  {journal} {\bibinfo  {journal} {Phys. Rev. Materials}\ }\textbf {\bibinfo
  {volume} {1}},\ \bibinfo {pages} {014002} (\bibinfo {year}
  {2017})}\BibitemShut {NoStop}%
\bibitem [{\citenamefont {Zhang}\ \emph {et~al.}(2012)\citenamefont {Zhang},
  \citenamefont {Liu}, \citenamefont {Duan}, \citenamefont {Liu},\ and\
  \citenamefont {Wu}}]{Sb:triv}%
  \BibitemOpen
  \bibfield  {author} {\bibinfo {author} {\bibfnamefont {P.}~\bibnamefont
  {Zhang}}, \bibinfo {author} {\bibfnamefont {Z.}~\bibnamefont {Liu}}, \bibinfo
  {author} {\bibfnamefont {W.}~\bibnamefont {Duan}}, \bibinfo {author}
  {\bibfnamefont {F.}~\bibnamefont {Liu}}, \ and\ \bibinfo {author}
  {\bibfnamefont {J.}~\bibnamefont {Wu}},\ }\href {\doibase
  10.1103/PhysRevB.85.201410} {\bibfield  {journal} {\bibinfo  {journal} {Phys.
  Rev. B}\ }\textbf {\bibinfo {volume} {85}},\ \bibinfo {pages} {201410}
  (\bibinfo {year} {2012})}\BibitemShut {NoStop}%
\bibitem [{\citenamefont {Bieniek}\ \emph {et~al.}(2016)\citenamefont
  {Bieniek}, \citenamefont {Wo{\'z}niak},\ and\ \citenamefont
  {Potasz}}]{BiAPPA}%
  \BibitemOpen
  \bibfield  {author} {\bibinfo {author} {\bibfnamefont {M.}~\bibnamefont
  {Bieniek}}, \bibinfo {author} {\bibfnamefont {T.}~\bibnamefont
  {Wo{\'z}niak}}, \ and\ \bibinfo {author} {\bibfnamefont {P.}~\bibnamefont
  {Potasz}},\ }\href@noop {} {\bibfield  {journal} {\bibinfo  {journal} {Acta
  Physica Polonica A}\ }\textbf {\bibinfo {volume} {2}},\ \bibinfo {pages}
  {609} (\bibinfo {year} {2016})}\BibitemShut {NoStop}%
\bibitem [{\citenamefont {Jin}\ and\ \citenamefont
  {Jhi}(2015)}]{jin2015quantum}%
  \BibitemOpen
  \bibfield  {author} {\bibinfo {author} {\bibfnamefont {K.-H.}\ \bibnamefont
  {Jin}}\ and\ \bibinfo {author} {\bibfnamefont {S.-H.}\ \bibnamefont {Jhi}},\
  }\href@noop {} {\bibfield  {journal} {\bibinfo  {journal} {Scientific
  reports}\ }\textbf {\bibinfo {volume} {5}} (\bibinfo {year}
  {2015})}\BibitemShut {NoStop}%
\bibitem [{\citenamefont {Chuang}\ \emph {et~al.}(2013)\citenamefont {Chuang},
  \citenamefont {Hsu}, \citenamefont {Chen}, \citenamefont {Huang},
  \citenamefont {Ozolins}, \citenamefont {Lin},\ and\ \citenamefont
  {Bansil}}]{Sb:nontriv}%
  \BibitemOpen
  \bibfield  {author} {\bibinfo {author} {\bibfnamefont {F.-C.}\ \bibnamefont
  {Chuang}}, \bibinfo {author} {\bibfnamefont {C.-H.}\ \bibnamefont {Hsu}},
  \bibinfo {author} {\bibfnamefont {C.-Y.}\ \bibnamefont {Chen}}, \bibinfo
  {author} {\bibfnamefont {Z.-Q.}\ \bibnamefont {Huang}}, \bibinfo {author}
  {\bibfnamefont {V.}~\bibnamefont {Ozolins}}, \bibinfo {author} {\bibfnamefont
  {H.}~\bibnamefont {Lin}}, \ and\ \bibinfo {author} {\bibfnamefont
  {A.}~\bibnamefont {Bansil}},\ }\href {\doibase 10.1063/1.4776734} {\bibfield
  {journal} {\bibinfo  {journal} {Applied Physics Letters}\ }\textbf {\bibinfo
  {volume} {102}},\ \bibinfo {pages} {022424} (\bibinfo {year}
  {2013})}\BibitemShut {NoStop}%
\bibitem [{\citenamefont {Wang}\ \emph {et~al.}(2014)\citenamefont {Wang},
  \citenamefont {Chen}, \citenamefont {Liu},\ and\ \citenamefont
  {Wang}}]{wang2014topological}%
  \BibitemOpen
  \bibfield  {author} {\bibinfo {author} {\bibfnamefont {D.}~\bibnamefont
  {Wang}}, \bibinfo {author} {\bibfnamefont {L.}~\bibnamefont {Chen}}, \bibinfo
  {author} {\bibfnamefont {H.}~\bibnamefont {Liu}}, \ and\ \bibinfo {author}
  {\bibfnamefont {X.}~\bibnamefont {Wang}},\ }\href@noop {} {\bibfield
  {journal} {\bibinfo  {journal} {EPL (Europhysics Letters)}\ }\textbf
  {\bibinfo {volume} {104}},\ \bibinfo {pages} {57011} (\bibinfo {year}
  {2014})}\BibitemShut {NoStop}%
\bibitem [{\citenamefont {Eisert}\ \emph {et~al.}(2010)\citenamefont {Eisert},
  \citenamefont {Cramer},\ and\ \citenamefont {Plenio}}]{Eisert:EE}%
  \BibitemOpen
  \bibfield  {author} {\bibinfo {author} {\bibfnamefont {J.}~\bibnamefont
  {Eisert}}, \bibinfo {author} {\bibfnamefont {M.}~\bibnamefont {Cramer}}, \
  and\ \bibinfo {author} {\bibfnamefont {M.~B.}\ \bibnamefont {Plenio}},\
  }\href {\doibase 10.1103/RevModPhys.82.277} {\bibfield  {journal} {\bibinfo
  {journal} {Rev. Mod. Phys.}\ }\textbf {\bibinfo {volume} {82}},\ \bibinfo
  {pages} {277} (\bibinfo {year} {2010})}\BibitemShut {NoStop}%
\bibitem [{\citenamefont {Amico}\ \emph {et~al.}(2008)\citenamefont {Amico},
  \citenamefont {Fazio}, \citenamefont {Osterloh},\ and\ \citenamefont
  {Vedral}}]{Ent:rev1}%
  \BibitemOpen
  \bibfield  {author} {\bibinfo {author} {\bibfnamefont {L.}~\bibnamefont
  {Amico}}, \bibinfo {author} {\bibfnamefont {R.}~\bibnamefont {Fazio}},
  \bibinfo {author} {\bibfnamefont {A.}~\bibnamefont {Osterloh}}, \ and\
  \bibinfo {author} {\bibfnamefont {V.}~\bibnamefont {Vedral}},\ }\href
  {\doibase 10.1103/RevModPhys.80.517} {\bibfield  {journal} {\bibinfo
  {journal} {Rev. Mod. Phys.}\ }\textbf {\bibinfo {volume} {80}},\ \bibinfo
  {pages} {517} (\bibinfo {year} {2008})}\BibitemShut {NoStop}%
\bibitem [{\citenamefont {Li}\ and\ \citenamefont
  {Haldane}(2008)}]{LiHaldane:ES}%
  \BibitemOpen
  \bibfield  {author} {\bibinfo {author} {\bibfnamefont {H.}~\bibnamefont
  {Li}}\ and\ \bibinfo {author} {\bibfnamefont {F.~D.~M.}\ \bibnamefont
  {Haldane}},\ }\href {\doibase 10.1103/PhysRevLett.101.010504} {\bibfield
  {journal} {\bibinfo  {journal} {Phys. Rev. Lett.}\ }\textbf {\bibinfo
  {volume} {101}},\ \bibinfo {pages} {010504} (\bibinfo {year}
  {2008})}\BibitemShut {NoStop}%
\bibitem [{\citenamefont {Oliveira}\ \emph {et~al.}(2014)\citenamefont
  {Oliveira}, \citenamefont {Ribeiro},\ and\ \citenamefont
  {Sacramento}}]{ENT:super1}%
  \BibitemOpen
  \bibfield  {author} {\bibinfo {author} {\bibfnamefont {T.~P.}\ \bibnamefont
  {Oliveira}}, \bibinfo {author} {\bibfnamefont {P.}~\bibnamefont {Ribeiro}}, \
  and\ \bibinfo {author} {\bibfnamefont {P.~D.}\ \bibnamefont {Sacramento}},\
  }\href {http://stacks.iop.org/0953-8984/26/i=42/a=425702} {\bibfield
  {journal} {\bibinfo  {journal} {Journal of Physics: Condensed Matter}\
  }\textbf {\bibinfo {volume} {26}},\ \bibinfo {pages} {425702} (\bibinfo
  {year} {2014})}\BibitemShut {NoStop}%
\bibitem [{\citenamefont {Fidkowski}(2010)}]{Fid:ES}%
  \BibitemOpen
  \bibfield  {author} {\bibinfo {author} {\bibfnamefont {L.}~\bibnamefont
  {Fidkowski}},\ }\href {\doibase 10.1103/PhysRevLett.104.130502} {\bibfield
  {journal} {\bibinfo  {journal} {Phys. Rev. Lett.}\ }\textbf {\bibinfo
  {volume} {104}},\ \bibinfo {pages} {130502} (\bibinfo {year}
  {2010})}\BibitemShut {NoStop}%
\bibitem [{\citenamefont {Fang}\ \emph {et~al.}(2013)\citenamefont {Fang},
  \citenamefont {Gilbert},\ and\ \citenamefont {Bernevig}}]{Bernevig:class}%
  \BibitemOpen
  \bibfield  {author} {\bibinfo {author} {\bibfnamefont {C.}~\bibnamefont
  {Fang}}, \bibinfo {author} {\bibfnamefont {M.~J.}\ \bibnamefont {Gilbert}}, \
  and\ \bibinfo {author} {\bibfnamefont {B.~A.}\ \bibnamefont {Bernevig}},\
  }\href {\doibase 10.1103/PhysRevB.87.035119} {\bibfield  {journal} {\bibinfo
  {journal} {Phys. Rev. B}\ }\textbf {\bibinfo {volume} {87}},\ \bibinfo
  {pages} {035119} (\bibinfo {year} {2013})}\BibitemShut {NoStop}%
\bibitem [{\citenamefont {Turner}\ \emph {et~al.}(2010)\citenamefont {Turner},
  \citenamefont {Zhang},\ and\ \citenamefont {Vishwanath}}]{Vish:inv}%
  \BibitemOpen
  \bibfield  {author} {\bibinfo {author} {\bibfnamefont {A.~M.}\ \bibnamefont
  {Turner}}, \bibinfo {author} {\bibfnamefont {Y.}~\bibnamefont {Zhang}}, \
  and\ \bibinfo {author} {\bibfnamefont {A.}~\bibnamefont {Vishwanath}},\
  }\href {\doibase 10.1103/PhysRevB.82.241102} {\bibfield  {journal} {\bibinfo
  {journal} {Phys. Rev. B}\ }\textbf {\bibinfo {volume} {82}},\ \bibinfo
  {pages} {241102} (\bibinfo {year} {2010})}\BibitemShut {NoStop}%
\bibitem [{\citenamefont {Hermanns}\ \emph {et~al.}(2014)\citenamefont
  {Hermanns}, \citenamefont {Salimi}, \citenamefont {Haque},\ and\
  \citenamefont {Fritz}}]{Fritz:ES}%
  \BibitemOpen
  \bibfield  {author} {\bibinfo {author} {\bibfnamefont {M.}~\bibnamefont
  {Hermanns}}, \bibinfo {author} {\bibfnamefont {Y.}~\bibnamefont {Salimi}},
  \bibinfo {author} {\bibfnamefont {M.}~\bibnamefont {Haque}}, \ and\ \bibinfo
  {author} {\bibfnamefont {L.}~\bibnamefont {Fritz}},\ }\href
  {http://stacks.iop.org/1742-5468/2014/i=10/a=P10030} {\bibfield  {journal}
  {\bibinfo  {journal} {Journal of Statistical Mechanics: Theory and
  Experiment}\ }\textbf {\bibinfo {volume} {2014}},\ \bibinfo {pages} {P10030}
  (\bibinfo {year} {2014})}\BibitemShut {NoStop}%
\bibitem [{\citenamefont {Kargarian}\ and\ \citenamefont
  {Fiete}(2010)}]{PhysRevB.82.085106}%
  \BibitemOpen
  \bibfield  {author} {\bibinfo {author} {\bibfnamefont {M.}~\bibnamefont
  {Kargarian}}\ and\ \bibinfo {author} {\bibfnamefont {G.~A.}\ \bibnamefont
  {Fiete}},\ }\href {\doibase 10.1103/PhysRevB.82.085106} {\bibfield  {journal}
  {\bibinfo  {journal} {Phys. Rev. B}\ }\textbf {\bibinfo {volume} {82}},\
  \bibinfo {pages} {085106} (\bibinfo {year} {2010})}\BibitemShut {NoStop}%
\bibitem [{\citenamefont {Alexandradinata}\ \emph {et~al.}(2011)\citenamefont
  {Alexandradinata}, \citenamefont {Hughes},\ and\ \citenamefont
  {Bernevig}}]{Alex:CM}%
  \BibitemOpen
  \bibfield  {author} {\bibinfo {author} {\bibfnamefont {A.}~\bibnamefont
  {Alexandradinata}}, \bibinfo {author} {\bibfnamefont {T.~L.}\ \bibnamefont
  {Hughes}}, \ and\ \bibinfo {author} {\bibfnamefont {B.~A.}\ \bibnamefont
  {Bernevig}},\ }\href {\doibase 10.1103/PhysRevB.84.195103} {\bibfield
  {journal} {\bibinfo  {journal} {Phys. Rev. B}\ }\textbf {\bibinfo {volume}
  {84}},\ \bibinfo {pages} {195103} (\bibinfo {year} {2011})}\BibitemShut
  {NoStop}%
\bibitem [{\citenamefont {Hughes}\ \emph {et~al.}(2011)\citenamefont {Hughes},
  \citenamefont {Prodan},\ and\ \citenamefont {Bernevig}}]{Hughes:inv}%
  \BibitemOpen
  \bibfield  {author} {\bibinfo {author} {\bibfnamefont {T.~L.}\ \bibnamefont
  {Hughes}}, \bibinfo {author} {\bibfnamefont {E.}~\bibnamefont {Prodan}}, \
  and\ \bibinfo {author} {\bibfnamefont {B.~A.}\ \bibnamefont {Bernevig}},\
  }\href {\doibase 10.1103/PhysRevB.83.245132} {\bibfield  {journal} {\bibinfo
  {journal} {Phys. Rev. B}\ }\textbf {\bibinfo {volume} {83}},\ \bibinfo
  {pages} {245132} (\bibinfo {year} {2011})}\BibitemShut {NoStop}%
\bibitem [{\citenamefont {Prodan}\ \emph {et~al.}(2010)\citenamefont {Prodan},
  \citenamefont {Hughes},\ and\ \citenamefont {Bernevig}}]{Bernevig:Disord}%
  \BibitemOpen
  \bibfield  {author} {\bibinfo {author} {\bibfnamefont {E.}~\bibnamefont
  {Prodan}}, \bibinfo {author} {\bibfnamefont {T.~L.}\ \bibnamefont {Hughes}},
  \ and\ \bibinfo {author} {\bibfnamefont {B.~A.}\ \bibnamefont {Bernevig}},\
  }\href {\doibase 10.1103/PhysRevLett.105.115501} {\bibfield  {journal}
  {\bibinfo  {journal} {Phys. Rev. Lett.}\ }\textbf {\bibinfo {volume} {105}},\
  \bibinfo {pages} {115501} (\bibinfo {year} {2010})}\BibitemShut {NoStop}%
\bibitem [{\citenamefont {Mondragon-Shem}\ \emph {et~al.}(2013)\citenamefont
  {Mondragon-Shem}, \citenamefont {Khan},\ and\ \citenamefont
  {Hughes}}]{Hughes:Disord1}%
  \BibitemOpen
  \bibfield  {author} {\bibinfo {author} {\bibfnamefont {I.}~\bibnamefont
  {Mondragon-Shem}}, \bibinfo {author} {\bibfnamefont {M.}~\bibnamefont
  {Khan}}, \ and\ \bibinfo {author} {\bibfnamefont {T.~L.}\ \bibnamefont
  {Hughes}},\ }\href {\doibase 10.1103/PhysRevLett.110.046806} {\bibfield
  {journal} {\bibinfo  {journal} {Phys. Rev. Lett.}\ }\textbf {\bibinfo
  {volume} {110}},\ \bibinfo {pages} {046806} (\bibinfo {year}
  {2013})}\BibitemShut {NoStop}%
\bibitem [{\citenamefont {Mondragon-Shem}\ and\ \citenamefont
  {Hughes}(2014)}]{Hughes:Disord2}%
  \BibitemOpen
  \bibfield  {author} {\bibinfo {author} {\bibfnamefont {I.}~\bibnamefont
  {Mondragon-Shem}}\ and\ \bibinfo {author} {\bibfnamefont {T.~L.}\
  \bibnamefont {Hughes}},\ }\href {\doibase 10.1103/PhysRevB.90.104204}
  {\bibfield  {journal} {\bibinfo  {journal} {Phys. Rev. B}\ }\textbf {\bibinfo
  {volume} {90}},\ \bibinfo {pages} {104204} (\bibinfo {year}
  {2014})}\BibitemShut {NoStop}%
\bibitem [{\citenamefont {Liu}\ and\ \citenamefont {Allen}(1995)}]{Liu:Allen}%
  \BibitemOpen
  \bibfield  {author} {\bibinfo {author} {\bibfnamefont {Y.}~\bibnamefont
  {Liu}}\ and\ \bibinfo {author} {\bibfnamefont {R.~E.}\ \bibnamefont
  {Allen}},\ }\href {\doibase 10.1103/PhysRevB.52.1566} {\bibfield  {journal}
  {\bibinfo  {journal} {Phys. Rev. B}\ }\textbf {\bibinfo {volume} {52}},\
  \bibinfo {pages} {1566} (\bibinfo {year} {1995})}\BibitemShut {NoStop}%
\bibitem [{\citenamefont {Slater}\ and\ \citenamefont
  {Koster}(1954)}]{Slater:Koster}%
  \BibitemOpen
  \bibfield  {author} {\bibinfo {author} {\bibfnamefont {J.~C.}\ \bibnamefont
  {Slater}}\ and\ \bibinfo {author} {\bibfnamefont {G.~F.}\ \bibnamefont
  {Koster}},\ }\href {\doibase 10.1103/PhysRev.94.1498} {\bibfield  {journal}
  {\bibinfo  {journal} {Phys. Rev.}\ }\textbf {\bibinfo {volume} {94}},\
  \bibinfo {pages} {1498} (\bibinfo {year} {1954})}\BibitemShut {NoStop}%
\bibitem [{\citenamefont {{Chadi}}(1977)}]{Chadi}%
  \BibitemOpen
  \bibfield  {author} {\bibinfo {author} {\bibfnamefont {D.~J.}\ \bibnamefont
  {{Chadi}}},\ }\href {\doibase 10.1103/PhysRevB.16.790} {\bibfield  {journal}
  {\bibinfo  {journal} {\prb}\ }\textbf {\bibinfo {volume} {16}},\ \bibinfo
  {pages} {790} (\bibinfo {year} {1977})}\BibitemShut {NoStop}%
\bibitem [{\citenamefont {Gonze}\ \emph {et~al.}(2016)\citenamefont {Gonze},
  \citenamefont {Jollet}, \citenamefont {Araujo}, \citenamefont {Adams},
  \citenamefont {Amadon}, \citenamefont {Applencourt}, \citenamefont {Audouze},
  \citenamefont {Beuken}, \citenamefont {Bieder}, \citenamefont {Bokhanchuk},
  \citenamefont {Bousquet}, \citenamefont {Bruneval}, \citenamefont {Caliste},
  \citenamefont {Côté}, \citenamefont {Dahm}, \citenamefont {Pieve},
  \citenamefont {Delaveau}, \citenamefont {Gennaro}, \citenamefont {Dorado},
  \citenamefont {Espejo}, \citenamefont {Geneste}, \citenamefont {Genovese},
  \citenamefont {Gerossier}, \citenamefont {Giantomassi}, \citenamefont
  {Gillet}, \citenamefont {Hamann}, \citenamefont {He}, \citenamefont {Jomard},
  \citenamefont {Janssen}, \citenamefont {Roux}, \citenamefont {Levitt},
  \citenamefont {Lherbier}, \citenamefont {Liu}, \citenamefont {Lukačević},
  \citenamefont {Martin}, \citenamefont {Martins}, \citenamefont {Oliveira},
  \citenamefont {Poncé}, \citenamefont {Pouillon}, \citenamefont {Rangel},
  \citenamefont {Rignanese}, \citenamefont {Romero}, \citenamefont {Rousseau},
  \citenamefont {Rubel}, \citenamefont {Shukri}, \citenamefont {Stankovski},
  \citenamefont {Torrent}, \citenamefont {Setten}, \citenamefont {Troeye},
  \citenamefont {Verstraete}, \citenamefont {Waroquiers}, \citenamefont
  {Wiktor}, \citenamefont {Xu}, \citenamefont {Zhou},\ and\ \citenamefont
  {Zwanziger}}]{DFT1}%
  \BibitemOpen
  \bibfield  {author} {\bibinfo {author} {\bibfnamefont {X.}~\bibnamefont
  {Gonze}}, \bibinfo {author} {\bibfnamefont {F.}~\bibnamefont {Jollet}},
  \bibinfo {author} {\bibfnamefont {F.~A.}\ \bibnamefont {Araujo}}, \bibinfo
  {author} {\bibfnamefont {D.}~\bibnamefont {Adams}}, \bibinfo {author}
  {\bibfnamefont {B.}~\bibnamefont {Amadon}}, \bibinfo {author} {\bibfnamefont
  {T.}~\bibnamefont {Applencourt}}, \bibinfo {author} {\bibfnamefont
  {C.}~\bibnamefont {Audouze}}, \bibinfo {author} {\bibfnamefont {J.-M.}\
  \bibnamefont {Beuken}}, \bibinfo {author} {\bibfnamefont {J.}~\bibnamefont
  {Bieder}}, \bibinfo {author} {\bibfnamefont {A.}~\bibnamefont {Bokhanchuk}},
  \bibinfo {author} {\bibfnamefont {E.}~\bibnamefont {Bousquet}}, \bibinfo
  {author} {\bibfnamefont {F.}~\bibnamefont {Bruneval}}, \bibinfo {author}
  {\bibfnamefont {D.}~\bibnamefont {Caliste}}, \bibinfo {author} {\bibfnamefont
  {M.}~\bibnamefont {Côté}}, \bibinfo {author} {\bibfnamefont
  {F.}~\bibnamefont {Dahm}}, \bibinfo {author} {\bibfnamefont {F.~D.}\
  \bibnamefont {Pieve}}, \bibinfo {author} {\bibfnamefont {M.}~\bibnamefont
  {Delaveau}}, \bibinfo {author} {\bibfnamefont {M.~D.}\ \bibnamefont
  {Gennaro}}, \bibinfo {author} {\bibfnamefont {B.}~\bibnamefont {Dorado}},
  \bibinfo {author} {\bibfnamefont {C.}~\bibnamefont {Espejo}}, \bibinfo
  {author} {\bibfnamefont {G.}~\bibnamefont {Geneste}}, \bibinfo {author}
  {\bibfnamefont {L.}~\bibnamefont {Genovese}}, \bibinfo {author}
  {\bibfnamefont {A.}~\bibnamefont {Gerossier}}, \bibinfo {author}
  {\bibfnamefont {M.}~\bibnamefont {Giantomassi}}, \bibinfo {author}
  {\bibfnamefont {Y.}~\bibnamefont {Gillet}}, \bibinfo {author} {\bibfnamefont
  {D.}~\bibnamefont {Hamann}}, \bibinfo {author} {\bibfnamefont
  {L.}~\bibnamefont {He}}, \bibinfo {author} {\bibfnamefont {G.}~\bibnamefont
  {Jomard}}, \bibinfo {author} {\bibfnamefont {J.~L.}\ \bibnamefont {Janssen}},
  \bibinfo {author} {\bibfnamefont {S.~L.}\ \bibnamefont {Roux}}, \bibinfo
  {author} {\bibfnamefont {A.}~\bibnamefont {Levitt}}, \bibinfo {author}
  {\bibfnamefont {A.}~\bibnamefont {Lherbier}}, \bibinfo {author}
  {\bibfnamefont {F.}~\bibnamefont {Liu}}, \bibinfo {author} {\bibfnamefont
  {I.}~\bibnamefont {Lukačević}}, \bibinfo {author} {\bibfnamefont
  {A.}~\bibnamefont {Martin}}, \bibinfo {author} {\bibfnamefont
  {C.}~\bibnamefont {Martins}}, \bibinfo {author} {\bibfnamefont
  {M.}~\bibnamefont {Oliveira}}, \bibinfo {author} {\bibfnamefont
  {S.}~\bibnamefont {Poncé}}, \bibinfo {author} {\bibfnamefont
  {Y.}~\bibnamefont {Pouillon}}, \bibinfo {author} {\bibfnamefont
  {T.}~\bibnamefont {Rangel}}, \bibinfo {author} {\bibfnamefont {G.-M.}\
  \bibnamefont {Rignanese}}, \bibinfo {author} {\bibfnamefont {A.}~\bibnamefont
  {Romero}}, \bibinfo {author} {\bibfnamefont {B.}~\bibnamefont {Rousseau}},
  \bibinfo {author} {\bibfnamefont {O.}~\bibnamefont {Rubel}}, \bibinfo
  {author} {\bibfnamefont {A.}~\bibnamefont {Shukri}}, \bibinfo {author}
  {\bibfnamefont {M.}~\bibnamefont {Stankovski}}, \bibinfo {author}
  {\bibfnamefont {M.}~\bibnamefont {Torrent}}, \bibinfo {author} {\bibfnamefont
  {M.~V.}\ \bibnamefont {Setten}}, \bibinfo {author} {\bibfnamefont {B.~V.}\
  \bibnamefont {Troeye}}, \bibinfo {author} {\bibfnamefont {M.}~\bibnamefont
  {Verstraete}}, \bibinfo {author} {\bibfnamefont {D.}~\bibnamefont
  {Waroquiers}}, \bibinfo {author} {\bibfnamefont {J.}~\bibnamefont {Wiktor}},
  \bibinfo {author} {\bibfnamefont {B.}~\bibnamefont {Xu}}, \bibinfo {author}
  {\bibfnamefont {A.}~\bibnamefont {Zhou}}, \ and\ \bibinfo {author}
  {\bibfnamefont {J.}~\bibnamefont {Zwanziger}},\ }\href {\doibase
  http://dx.doi.org/10.1016/j.cpc.2016.04.003} {\bibfield  {journal} {\bibinfo
  {journal} {Computer Physics Communications}\ }\textbf {\bibinfo {volume}
  {205}},\ \bibinfo {pages} {106 } (\bibinfo {year} {2016})}\BibitemShut
  {NoStop}%
\bibitem [{\citenamefont {Togo}\ and\ \citenamefont {Tanaka}(2015)}]{DFT2}%
  \BibitemOpen
  \bibfield  {author} {\bibinfo {author} {\bibfnamefont {A.}~\bibnamefont
  {Togo}}\ and\ \bibinfo {author} {\bibfnamefont {I.}~\bibnamefont {Tanaka}},\
  }\href {\doibase http://dx.doi.org/10.1016/j.scriptamat.2015.07.021}
  {\bibfield  {journal} {\bibinfo  {journal} {Scripta Materialia}\ }\textbf
  {\bibinfo {volume} {108}},\ \bibinfo {pages} {1 } (\bibinfo {year}
  {2015})}\BibitemShut {NoStop}%
\bibitem [{\citenamefont {Parlinski}\ \emph {et~al.}(1997)\citenamefont
  {Parlinski}, \citenamefont {Li},\ and\ \citenamefont {Kawazoe}}]{DFT3}%
  \BibitemOpen
  \bibfield  {author} {\bibinfo {author} {\bibfnamefont {K.}~\bibnamefont
  {Parlinski}}, \bibinfo {author} {\bibfnamefont {Z.~Q.}\ \bibnamefont {Li}}, \
  and\ \bibinfo {author} {\bibfnamefont {Y.}~\bibnamefont {Kawazoe}},\ }\href
  {\doibase 10.1103/PhysRevLett.78.4063} {\bibfield  {journal} {\bibinfo
  {journal} {Phys. Rev. Lett.}\ }\textbf {\bibinfo {volume} {78}},\ \bibinfo
  {pages} {4063} (\bibinfo {year} {1997})}\BibitemShut {NoStop}%
\bibitem [{\citenamefont {Evarestov}\ and\ \citenamefont {Losev}(2009)}]{DFT4}%
  \BibitemOpen
  \bibfield  {author} {\bibinfo {author} {\bibfnamefont {R.~A.}\ \bibnamefont
  {Evarestov}}\ and\ \bibinfo {author} {\bibfnamefont {M.~V.}\ \bibnamefont
  {Losev}},\ }\href {\doibase 10.1002/jcc.21259} {\bibfield  {journal}
  {\bibinfo  {journal} {Journal of Computational Chemistry}\ }\textbf {\bibinfo
  {volume} {30}},\ \bibinfo {pages} {2645} (\bibinfo {year}
  {2009})}\BibitemShut {NoStop}%
\bibitem [{\citenamefont {Singh}\ and\ \citenamefont
  {Romero}(2017)}]{DFT:BiSb}%
  \BibitemOpen
  \bibfield  {author} {\bibinfo {author} {\bibfnamefont {S.}~\bibnamefont
  {Singh}}\ and\ \bibinfo {author} {\bibfnamefont {A.~H.}\ \bibnamefont
  {Romero}},\ }\href {\doibase 10.1103/PhysRevB.95.165444} {\bibfield
  {journal} {\bibinfo  {journal} {Phys. Rev. B}\ }\textbf {\bibinfo {volume}
  {95}},\ \bibinfo {pages} {165444} (\bibinfo {year} {2017})}\BibitemShut
  {NoStop}%
\bibitem [{\citenamefont {Plenio}\ and\ \citenamefont
  {Virmani}(2007)}]{Plenio:2007zz}%
  \BibitemOpen
  \bibfield  {author} {\bibinfo {author} {\bibfnamefont {M.~B.}\ \bibnamefont
  {Plenio}}\ and\ \bibinfo {author} {\bibfnamefont {S.}~\bibnamefont
  {Virmani}},\ }\href@noop {} {\bibfield  {journal} {\bibinfo  {journal}
  {Quant. Inf. Comput.}\ }\textbf {\bibinfo {volume} {7}},\ \bibinfo {pages}
  {1} (\bibinfo {year} {2007})}\BibitemShut {NoStop}%
\bibitem [{\citenamefont {Peschel}(2003)}]{Peschel}%
  \BibitemOpen
  \bibfield  {author} {\bibinfo {author} {\bibfnamefont {I.}~\bibnamefont
  {Peschel}},\ }\href@noop {} {\bibfield  {journal} {\bibinfo  {journal}
  {Journal of Physics A: Mathematical and General}\ }\textbf {\bibinfo {volume}
  {36}},\ \bibinfo {pages} {L205} (\bibinfo {year} {2003})}\BibitemShut
  {NoStop}%
\bibitem [{\citenamefont {Ryu}\ and\ \citenamefont {Hatsugai}(2006)}]{Ryu:EE}%
  \BibitemOpen
  \bibfield  {author} {\bibinfo {author} {\bibfnamefont {S.}~\bibnamefont
  {Ryu}}\ and\ \bibinfo {author} {\bibfnamefont {Y.}~\bibnamefont {Hatsugai}},\
  }\href {\doibase 10.1103/PhysRevB.73.245115} {\bibfield  {journal} {\bibinfo
  {journal} {Phys. Rev. B}\ }\textbf {\bibinfo {volume} {73}},\ \bibinfo
  {pages} {245115} (\bibinfo {year} {2006})}\BibitemShut {NoStop}%
\bibitem [{\citenamefont {Chandran}\ \emph {et~al.}(2014)\citenamefont
  {Chandran}, \citenamefont {Khemani},\ and\ \citenamefont
  {Sondhi}}]{Sondhi:univ}%
  \BibitemOpen
  \bibfield  {author} {\bibinfo {author} {\bibfnamefont {A.}~\bibnamefont
  {Chandran}}, \bibinfo {author} {\bibfnamefont {V.}~\bibnamefont {Khemani}}, \
  and\ \bibinfo {author} {\bibfnamefont {S.~L.}\ \bibnamefont {Sondhi}},\
  }\href {\doibase 10.1103/PhysRevLett.113.060501} {\bibfield  {journal}
  {\bibinfo  {journal} {Phys. Rev. Lett.}\ }\textbf {\bibinfo {volume} {113}},\
  \bibinfo {pages} {060501} (\bibinfo {year} {2014})}\BibitemShut {NoStop}%
\bibitem [{\citenamefont {Kane}\ and\ \citenamefont
  {Mele}(2005{\natexlab{b}})}]{KaneMele:QSHE2}%
  \BibitemOpen
  \bibfield  {author} {\bibinfo {author} {\bibfnamefont {C.~L.}\ \bibnamefont
  {Kane}}\ and\ \bibinfo {author} {\bibfnamefont {E.~J.}\ \bibnamefont
  {Mele}},\ }\href {\doibase 10.1103/PhysRevLett.95.226801} {\bibfield
  {journal} {\bibinfo  {journal} {Phys. Rev. Lett.}\ }\textbf {\bibinfo
  {volume} {95}},\ \bibinfo {pages} {226801} (\bibinfo {year}
  {2005}{\natexlab{b}})}\BibitemShut {NoStop}%
\bibitem [{\citenamefont {Harrison}(2012)}]{harrison}%
  \BibitemOpen
  \bibfield  {author} {\bibinfo {author} {\bibfnamefont {W.~A.}\ \bibnamefont
  {Harrison}},\ }\href@noop {} {\emph {\bibinfo {title} {Electronic structure
  and the properties of solids: the physics of the chemical bond}}}\ (\bibinfo
  {publisher} {Courier Corporation},\ \bibinfo {year} {2012})\BibitemShut
  {NoStop}%
\end{thebibliography}%

\end{document}